\newif\ifref
\reffalse          

\ifref
  \documentstyle[graphics,referee,endfloat]{mn}
  \renewcommand{\efloatseparator}{}
\else
  \documentstyle[graphics]{mn}
\fi
\title[Linearizing the Observed Power Spectrum]
      {Linearizing the Observed Power Spectrum}
\author[Clay. C. Smith et al.]
       {Clay~C.~Smith,$^1$\footnotemark[1]
        Anatoly~Klypin,$^1$\footnotemark[1]
        Michael~A.~K.~Gross,$^2$\footnotemark[1]
        Joel~R.~Primack,$^2$\footnotemark[1]
        \newauthor
        and Jon~Holtzman$^1$\footnotemark[1]\\
        $^1$Department of Astronomy, New Mexico State University, Las Cruces, NM 88003 USA\\
        $^2$Physics Department, University of California, Santa Cruz, CA 95064 USA\\
       }
\newlength{\figwidth}
\ifref
\else
\fi

\ifCUPmtlplainloaded \else
  \DeclareSymbolFont{UPM}{U}{eur}{m}{n}
  \SetSymbolFont{UPM}{bold}{U}{eur}{b}{n}
  \DeclareSymbolFont{AMSa}{U}{msa}{m}{n}
  \DeclareMathSymbol{\upi}{0}{UPM}{"19}
  \DeclareMathSymbol{\umu}{0}{UPM}{"16}
  \DeclareMathSymbol{\upartial}{0}{UPM}{"40}
  \DeclareMathSymbol{\leqslant}{3}{AMSa}{"36}
  \DeclareMathSymbol{\geqslant}{3}{AMSa}{"3E}
\fi

\newcommand{\etal}{et al.}      
\newcommand{\lcdm}{\mbox{$\Lambda$CDM}}
\newcommand{\kms}{\mbox{km s$^{-1}$}}
\newcommand{\kmsmpc}{\mbox{km s$^{-1}$ Mpc$^{-1}$}}
\newcommand{\hmpcinv}{\mbox{$h$ Mpc$^{-1}$}}
\newcommand{\hmpc}{\mbox{$h^{-1}$ Mpc}}
\newcommand{\iras}{\emph{IRAS}}
\newcommand{\cobe}{\emph{COBE}}
\newcommand{\potent}{\textsc{potent}}

\begin{document}

\maketitle
\begin{abstract}
Semi-analytic treatment of the power spectrum with the approximation
of constant linear bias provides a way to compare cosmological models
to a large amount of data, as Peacock \& Dodds (1994, 1996; hereafter
PD94 and PD96, respectively)
have shown. By applying the appropriate corrections to the
observational power spectrum it is possible to recover the underlying
linear power spectrum for any given cosmological model.  Using
extensive N-body simulations we carefully test and calibrate all
important corrections. To demonstrate that the method is applicable to
a wide range of cosmological models, we tested the standard $\Omega=1$
cold dark matter (CDM) model, $\Omega<1$ models that include a
cosmological constant (\lcdm), and $\Omega=1$ models with a mixture of
cold and hot dark matter (CHDM), both with one massive neutrino and two equal
mass neutrinos. The \lcdm\ and CHDM models are
normalized to \cobe\ and to cluster abundances.  Our tests indicate that
the improved linear--nonlinear mapping recently suggested by PD96
for treating CDM-type power spectra works well for a wide range of
scale-dependent power spectra.  However, we find that the recovery of
the linear power spectrum from observations following PD94,
which is often used to test cosmological models, can be misleading
because the corrections are model-dependent.  A model should not be
judged based on the linear spectrum recovered by that procedure, but
must be compared with the recovered spectrum for that particular
model. When we apply the proper corrections for a given model to the
observational power spectrum, we find that no model in our test group
recovers the linear power spectrum well for the bias values suggested
by PD94 between Abell, radio, optical, and \iras\ catalogs:
$b_{\rm A}:b_{\rm R}:b_{\rm O}:b_{\rm I}=4.5:1.9:1.3:1.0$, with $b_{\rm I}=1.0$.
The recovered linear \lcdm\ and CHDM power spectra were systematically
below their respective linear power spectra using $b_{\rm I}=1.0$.  When we
allow $b_{\rm I}$ to vary (keeping the same bias ratios) we find that: (i) CHDM
models give very good fits to observations if optically-selected
galaxies are slightly biased ($b_{\rm O}\sim 1.1$). (ii) Most \lcdm\ models
give worse but acceptable fits if blue galaxies are considerably
antibiased ($0.6\la b_{\rm O} \la 0.9$) and fail if optical galaxies are
biased. (iii) There is a universal shape of the recovered linear power
spectrum of all \lcdm\ models over their entire range of explored
wavenumbers, $0.01\la k\la 0.6$ \hmpcinv. For a given bias,
recovered linear power spectra of CDM and CHDM models are nearly the
same as that of \lcdm\ in the region $0.01\la k\la 0.2$
\hmpcinv\ but diverge from this spectrum at higher $k$.
We tabulate the recovered linear spectra, and also the initial linear 
spectra, for all the models considered.
\end{abstract}

\begin{keywords}
cosmology: theory -- large--scale structure of Universe
\end{keywords}

\footnotetext[1]{E-mail:
                 \ifref \else \\ \fi
                 aklypin@nmsu.edu~(AK);
                 gross@physics.ucsc.edu~(MAKG);
                 \ifref \else \\ \fi
                 joel@ucolick.org~(JRP);
                 holtz@nmsu.edu~(JH)}

\renewcommand{\thefootnote}{{\arabic{footnote}}}

\section{Introduction}

The power spectrum of density fluctuations $P(k)$ is a powerful tool
for investigating the large-scale structure of the Universe.  It is a
useful statistical test of the distribution of matter from the scales
of galaxy groups through scales larger than superclusters, and it
discriminates between cosmological models (see, for example, Liddle et
al. 1996ab; Ma 1996; Klypin, Primack, \& Holtzman 1996, hereafter
KPH96).
However, there are several obstacles that must be overcome to relate
the observed nonlinear power spectrum to theoretical linear power
spectra. Because observations typically give estimates for luminous
objects in redshift space while theoretical models describe dark
matter in real space, the differences between luminous and dark
matter must be taken into account before a direct comparison between
data and theory can be made.

Since the radial positions of galaxies are determined from their
observed recession velocities, any peculiar motions (from either
linear or nonlinear growth of fluctuations) will distort their
placement.  Kaiser (1987)
estimated the effect for long waves in the linear regime and included
a scale-independent bias. This correction applies to the region
$k\la 0.1$ \hmpcinv\ but not to smaller scales which are in the
nonlinear regime.  To treat this region, Peacock (1992)
used a simplified model to estimate the effect of random motions in
collapsed objects on the observed power spectrum.  Peacock \& Dodds
(1994, hereafter PD94) combined the two separate effects, along with
the overall bias level, into a single convenient expression that is
claimed to remove these effects from the nonlinear power spectrum.

Additional nonlinear effects must be accounted for as well. Theory
provides a linear power spectrum for a cosmological model, but the
spectrum of the real universe has undergone significant modification
because of nonlinear evolution.  Numerical simulations are useful for
estimating the nonlinear power spectrum but are time consuming and
computationally expensive.  A more practical approach would be to
devise a general method for analytically constructing the nonlinear
power spectrum from the linear power spectrum.
This would allow any model to be evaluated quickly and open the
possibility of directly removing nonlinear effects from a real power
spectrum.  Throughout this paper we will refer to the method of
recovering the linear spectrum from the raw observational spectrum by
applying a set of corrections as the `linearization' procedure.  We
will refer to the inverse of linearization, the method of predicting
the nonlinear spectrum of galaxies starting from the linear dark
matter spectrum, as the `nonlinearization' procedure.

Several advances in developing linearization methods have been made in
the last few years. By considering the collapse of spherical
structures and the nonlinear evolution in the limit of stable
clustering, Hamilton \etal\ (1991, hereafter HKLM) devised an
analytical method for relating the linear correlation function on a
linear scale to the nonlinear correlations on a smaller, nonlinear
scale.  PD94
extended this method to power spectra, allowed for open models and
models with a cosmological constant, and provided a fitting formula
optimized by a series of numerical simulations.  This semi-analytical
method worked reasonably well for scale-free power-law spectra, but
failed for models with spectral indices $n\la -2$ including
scale-dependent spectra like those of CDM models that tend to $n\sim
-3$ on the smallest scales.  Like HKLM, PD94 expected that their
series of corrections would be nearly model-independent and could be
applied to observational power spectra to arrive at the underlying
linear power spectrum of the Universe. Not long after, however, Jain,
Mo, \& White (1995) found that there is indeed a significant
dependence on the model through the spectral index $n$ of the linear
power spectrum and suggested that an `effective index' be used to
treat curving spectra like those of CDM or CHDM models. This prompted
Peacock \& Dodds (1996, hereafter PD96) to revise their earlier
fitting formula to include dependence on $n$, and propose that $n$
should vary with scale to treat curving spectra.

In this paper we adopt the analytical form of the redshift corrections
of PD94 and the improved linear-nonlinear semi-analytical
approximation of PD96, although we find it necessary to modify both of
these slightly.  We have extensively tested each step of the procedure
using numerical simulations of unbiased dark matter in CDM, \lcdm, and
CHDM models.  We find that the redshift corrections of PD94 are close
to those that we determine from our simulations, although they are
improved when we use the appropriate 1D velocity dispersions for each
model instead of universally applying 300 \kms.  We also demonstrate
that our modification  of the linear-nonlinear mapping of PD96 works
very well for the cosmologies that we examine; CDM, \lcdm, and CHDM
models are all well fit.

Having found that the method can handle a variety of models, we
applied it to a compilation of observational power spectra (APM,
APM/Stromlo, CfA-2, QDOT, and radio galaxies, and Abell clusters). Our
intended goal is to demonstrate that the linear power spectrum can be
recovered from concrete catalogs of objects in redshift space or real
space for a number of interesting cosmological models and then use
this technique to determine which model best fits the observed power
spectrum. Because the corrections are model-dependent (mostly through
the shape of the linear power spectrum), no simple scheme can
reproduce the underlying linear spectrum without detailed information
about the model.  This dictates that we must select well-defined
models and try to reconstruct their linear power spectra using the
appropriate corrections {\em for each model}.  For this reason, the
widely used power spectrum of PD94 is of limited utility, since no
model was specified. Several papers have used this power spectrum as a
means for testing cosmological models (e.g., Liddle et al. 1996ab; Ma
1996; Coble, Dodelson, \& Frieman 1996).  We assert that this is
incorrect and can lead to incorrect conclusions. The only measure of
success of a particular model in this context is its ability to
reconstruct the linear
power spectrum from the \emph{model-dependent} linearization of the
observational power spectrum.

We have followed PD94 in assuming that the bias is linear and
constant, and in our choice of observational data, but we have
propagated the errors somewhat differently.  We also considered a form
of scale-dependent bias for one model.  We have used the bias ratios
of PD94, but we allowed the reference level $b_{\rm I}$ to float and by
adjusting the overall bias level we optimized the fit of the
linearization to the linear power spectrum for each model. We find
that the linearization of the observational power spectra according to
the \lcdm\ family of models can all be described by the same shape and
amplitude for a given bias level.  In contrast, linearizations of CHDM
models do not have a common form.  Moreover, it appears that there is
no universal linearized power spectrum since CDM, \lcdm, and CHDM
linearizations are all easily distinguishable. In addition, we arrive
at the conclusion that optically selected galaxies must be antibiased
($b_{\rm O}\simeq 0.6-0.9$) if they are to be compatible with the \lcdm\
family of models.  The estimates of the optical biases of different
models from \potent\ analyses by Hudson \etal\ (1995)
seem to indicate that a small amount of antibias is acceptable for
some models, though Peacock (1996) would disagree.  Optical galaxies
in CHDM models must be slightly biased ($b_{\rm O}\simeq 1.1$), which would
probably tend to favor CHDM models.

This paper is organized as follows.  The details of our simulations
are described in \S\ref{numsims}.  \S\ref{obsdata} lists
the various power spectra in the observational data set. In \S\ref{method}
we review the semi-analytic method in depth.  We discuss
the tests of the method and its application to our simulations in
\S\ref{appsims}.  The linearized observational power spectra for
all the models that we consider are presented in \S\ref{appdata}.
Our results are summarized in \S\ref{conclusions}.  Appendix~\ref{app:fits}
gives theoretical power spectra for all the models considered.

\section{Numerical simulations}
\label{numsims}

In order to fully test our semi-analytical approximations, we
performed an array of numerical simulations of several different
cosmological models.  All simulations were performed using standard
Particle-Mesh (PM) codes (Hockney \& Eastwood 1981; Kates, Kotok, \&
Klypin 1991; Smith 1995; Klypin, Nolthenius, \& Primack 1997; Gross
\etal\ 1997).
Table~\ref{modelparam_table}
\begin{table*}
\begin{minipage}{320pt}
\begin{center}
\caption {Summary of model parameters}
\label{modelparam_table}
\begin{tabular*}{320pt}{lccccccccc}
Model & $\Omega_0$$^a$ & $\Omega_\Lambda$$^b$ &
$\Omega_{\rm c+b}$$^c$ & $\Omega_{\rm b}$$^d$ & $h$$^e$ & $n$$^f$ &
$\sigma_8$$^g$ & Q$^h$ & $\Gamma$$^i$ \\
\hline
SCDM & 1.0 & 0.0 & 1.0 & 0.00 & 0.5 & 1.00 & 0.67 & 8.49 & 0.50 \\
\\
\lcdm$_{0.3}$ & 0.3 & 0.7 & 0.3 & 0.026 & 0.7 & 1.00 & 1.10 & 22.0 & 0.20 \\
\lcdm$_{0.4}$ & 0.4 & 0.6 & 0.4 & 0.035 & 0.6 & 0.975 & 1.00 & 21.8 & 0.23 \\
T\lcdm$_{0.4}$ & 0.4 & 0.6 & 0.4 & 0.035 & 0.6 & 0.90 & 0.873 & 24.7 & 0.23 \\
\lcdm$_{0.5}$ & 0.5 & 0.5 & 0.5 & 0.035 & 0.6 & 1.00 & 0.818 & 13.93 & 0.29 \\
\\
CHDM$_{0.7}$ & 1.0 & 0.0 & 0.7 & 0.10 & 0.5 & 1.00 & 0.676 & 17.0 & 0.39 \\
CHDM$_{0.8}$ & 1.0 & 0.0 & 0.8 & 0.10 & 0.5 & 1.00 & 0.719 & 18.44 & 0.34\\
\hline
\end{tabular*}
\end{center}

$^a$ Matter density parameter.

$^b$ Cosmological constant density parameter $\Omega_\Lambda \equiv
\Lambda/(3H_0^2)$.

$^c$ Cold dark matter density parameter + baryon matter density
parameter.

$^d$ Baryonic matter density parameter.

$^e$ The Hubble parameter specified as $h\equiv H_0 / (100$ \kmsmpc$)$.

$^f$ Index of the power spectrum on large scales.

$^g$ Linear mass fluctuation amplitude $(\Delta M/M)_{\rm rms}$
smoothed with a top-hat filter of radius 8 \hmpc.

$^h$ Quadrupole of CMBR anisotropy Q$_{\rm RMS-ps}$.

$^i$ Sugiyama (1995) shape parameter $\Gamma=\Omega_0 h \exp(-\Omega_{\rm b}
- \Omega_{\rm b}/\Omega_0)$.  The actual spectral shape of the CHDM models
is steeper for large $k$ than the \lcdm\ models, contrary to the
impression that the high values of $\Gamma$ for these models might
give. Note that the value of the modified shape parameter
$\Gamma=0.5(0.95\sigma_8/3.5 \sigma_{25})^{3.33})$ introduced in
Borgani \etal\ (1997) is nearly identical to the Sugiyama shape
parameter for all but the two CHDM models, for which it would be 0.16
and 0.19, respectively.
\end{minipage}
\end{table*}
lists the models and gives their defining
parameters.  We used the `standard' cold dark matter model (SCDM)
with bias $b=\sigma_8^{-1}=1.5$, four different flat \lcdm\ models,
and two $\Omega=1$ CHDM models.
The CHDM$_{0.7}$ model has one neutrino species and the CHDM$_{0.8}$
model has two equally massive neutrino species.  Except for SCDM, all
our simulations started from \cobe-normalized Gaussian initial
conditions.

Table~\ref{simsum_table}
\begin{table*}
\begin{minipage}{301pt}
\begin{center}
\caption {Summary of simulations used in this paper}
\label{simsum_table}
\begin{tabular*}{301pt}{lccccc}
Model & mesh$^a$ & particles$^b$ & box size$^c$ & resolution$^d$
& realizations$^e$ \\
& size & & (\hmpc) & ($h^{-1}$ kpc) & \\
\hline
SCDM & $768^3$ & $384^3$ & 300 & 390 & 1 \\
\\
$\Lambda $CDM$_{0.3}$ & $512^3$ & $256^3$ & 200 & 390 & 3 \\
$\Lambda $CDM$_{0.4}$ & $512^3$ & $256^3$ & 200 & 390 & 3 \\
T$\Lambda $CDM$_{0.4}$ & $768^3$ & $384^3$ & 300 & 390 & 1 \\
T$\Lambda $CDM$_{0.4}$ & $512^3$ & $256^3$ & 100 & 195 & 3 \\
T$\Lambda $CDM$_{0.4}$ & $512^3$ & $256^3$ & 50 & 98 & 3 \\
$\Lambda $CDM$_{0.5}$ & $768^3$ & $384^3$ & 300 & 390 & 1 \\
\\
CHDM$_{0.7}$ & $768^3$ & $3\times 256^3$ & 255 & 332 & 1 \\
CHDM$_{0.8}$ & $768^3$ & $3\times 384^3$ & 300 & 390 & 1 \\
\hline
\end{tabular*}
\end{center}

$^a$ Total number of PM divisions.

$^b$ Number of particles in the box.

$^c$ Total size of the box.

$^d$ Box size divided by the number of PM divisions.

$^e$ Number of simulations with different realizations that were
averaged.
\end{minipage}
\end{table*}
shows the box sizes, resolutions and number
of realizations for each model.  Multiple box sizes and realizations
were simulated for many models in order to reduce the impact of two
effects which can change estimates of the nonlinear power spectrum as
calculated from simulations.  (1) Cosmic variance can cause
significant fluctuation of the power spectrum on large scales due to
poor counting statistics.  Averaging the power spectra of several
smaller realizations together reduces this effect significantly. (2)
Small box sizes ($L\la 100$ \hmpc) have the effect of reducing
the power in the nonlinear region of the power spectrum by as much as
50 per cent, leading to an underestimation of the scale at which nonlinear
effects begin to appear.  This is due to the fact that long waves
(which are coupled to shorter waves) cannot be present in the box.
Boxes with $L\ga 200$ \hmpc\ are better for
constructing accurate power spectra of the quasi-nonlinear region
because the first several bins at small $k$ then lie in the truly linear
regime.  By using boxes with $L=300$ \hmpc\ for most of the
models discussed in this paper, we address both problems.

For CDM-type spectra like those that we examine in this paper, (with
or without a cosmological constant, but not including CHDM), the
primordial spectrum is modified by physical processes that can be
expressed in a scale-dependent transfer function, $T(k)$. Since the
shape of the CDM spectrum does not change much after the epoch of
equality, it is a function of $k$ only. The power spectrum of the
tilted CDM-type models considered here can thus be approximated as
\begin{equation}
P(k)=A\,k^n\,T^2(k)\frac{g^2(\Omega(t))}{g^2(\Omega_0)}
\label{PS_CDMgen}
\end{equation}
where $g(\Omega,\Omega_\Lambda)$ is the growth rate of fluctuations used
to specify the power spectrum at different epochs and $T(k)$ is the transfer
function.  We use the approximation of Carroll, Press, \& Turner
(1992) (cf. also Lahav et al. 1991),
\begin{equation}
g(\Omega,\Omega_\Lambda)=\frac 52\Omega\left[ \Omega^{4/7}-\Omega_\Lambda
+(1+\frac {\Omega}2)(1+\frac{\Omega_\Lambda }{70})\right] ^{-1} ,
\label{gOmdef}
\end{equation}
where $\Omega_\Lambda \equiv \Lambda/(3H_0^2)$.  The SCDM transfer
function that we used was the commonly used approximation of Bardeen
\etal\ (1986, BBKS),
which has the form
\begin{equation}
\ifref \else \begin{array}{lr}\lefteqn{ \fi        
T(q)=\frac{\ln (1+2.34q)}{2.34q}
\ifref \else }&\\&\times \fi                       
\left[ 1+3.89q+(16.1q)^2 +(5.46q)^3 +(6.71q)^4 \right]^{-1/4}
\ifref \else \end{array} \fi                       
\label{BBKSfit}
\end{equation}
where $q=k/(h\Gamma)$ and $\Gamma$ is the `shape parameter' defined
by Sugiyama (1995) (but cf. Hu \& Sugiyama 1996 for improved spectra).
\begin{equation}
\Gamma=\Omega_0 h \exp(-\Omega_{\rm b} - \Omega_{\rm b}/\Omega_0) .
\label{Gammadef}
\end{equation}
Here, $\Omega_{\rm b}$ is the contribution of baryonic matter to the
cosmological density.  However, we only used the BBKS transfer
function with $\Gamma=0.5$ for SCDM.  For our other simulations, the
transfer functions of the \lcdm\ and CHDM models we used were
calculated from a full Boltzmann treatment updating that in Holtzman
(1989); fitting formulas are given in Appendix~\ref{app:fits}.
For CHDM models, the different time-dependent growth rates of the cold
and hot species means that the shape of the power spectrum changes
over time.  Incorporating the growth rates into the transfer function
gives a general form of the CHDM power spectrum,
\begin{equation}
P_{\rm CHDM}(k)=A\,k^n\,T^2(k,\Omega(t)) .
\label{PS_CHDMgen}
\end{equation}
The CHDM transfer functions were essentially those of Klypin \etal\ 
(1993) and Primack \etal\ (1995); cf. Appendix~\ref{app:fits}.

In all, we have three classes of dark matter models with a total of
seven distinct models.  Our sampling covers a wide range of power
spectra with shallow (SCDM), medium (\lcdm), and steep (CHDM) spectral
indices on small scales, which allows us to test our method
thoroughly.

\section{Observational data}
\label{obsdata}

The observational power spectra to which we will apply our
linearization methods consist of six independent data sets and two
cross correlations between data sets, as assembled by PD94.  Included
in the sample are catalogs of optical clusters, radio galaxies,
optical galaxies, and galaxies from the \iras\ catalog.  Each is briefly
described below.

One set of a real-space power spectrum is included, the APM power
spectrum (Baugh \& Efstathiou 1993).
It is arrived at by deprojecting the angular clustering of individual
galaxies from the APM survey.  Because it only deals with angular
positions, it directly yields the real-space power spectrum without
the need to correct for redshift-space effects. Of the full data set,
we choose to examine only points in the region
$0.015<k<1$ \hmpcinv. Power spectra from three
catalogs of galaxies in redshift-space were used: the Stromlo/APM
survey (Loveday \etal\ 1992);
the CfA survey (Vogeley \etal\ 1992);
and \iras\ galaxies from the QDOT sample (Feldman \etal\ 1994).
A straight mean of the two power spectra in the CfA paper was taken.
Two catalogs of clusters of galaxies in redshift-space were used, the
power spectrum of Abell clusters of richness class 1 or higher
(Peacock \& West 1992)
and the power spectrum of radio galaxies from (Peacock \& Nocholson
1991)
The two cross-correlations are between \iras\ galaxies and Abell
clusters and between \iras\ galaxies and radio galaxies (Mo \etal\ 1993).

Different biases should be assigned to the different data sets based
on how they were selected.  PD94 defined four adjustable bias
parameters for the four categories in this sample: $b_{\rm A}$ for Abell
clusters, $b_{\rm R}$ for radio galaxies, $b_{\rm O}$ for optically-selected
galaxies, and $b_{\rm I}$ for \iras\ galaxies.  Using a likelihood
method, PD94 chose the `best' ratio of these bias parameters to be
\begin{equation}
b_{\rm A}:b_{\rm R}:b_{\rm O}:b_{\rm I}=4.5:1.9:1.3:1.0 ,
\label{biasratio}
\end{equation}
normalized to $b_{\rm I}=1$.  In this paper we adopt the above ratio
but use $b_{\rm I}$ as a reference level which can be adjusted. The
necessity of this step will be discussed in \S\ref{appdata}.

This sample well represents the observable power spectrum in the
quasi-nonlinear regime up to such small $k$ that there should be no
apparent nonlinear effects.  The method that is described in the
following section performs very well in this region of the power
spectrum. Using this method we will attempt to remove all of the
observational effects from these power spectra.

\section{Method}
\label{method}

The linearization method that we use is wholly based on that of PD94
and the later modification by PD96 which improved the handling of
CDM-type power spectra. In order to recover a pure linear spectrum
from real observations three effects must be corrected for: redshift
distortions, bias and nonlinear growth.  In practice, corrections for
the redshift distortions and bias are inseparable and are applied
together through a single equation (\S 4.1).
The resulting nonlinear spectrum can be mapped to the linear spectrum
on the linear scale.

The contribution to the fractional density variance per bin of $\ln
k$, denoted $\Delta^2(k)$ (Peebles 1980),
is related to the power spectrum $P(k)=\left< \left| \delta_{\bf k}
\right| ^2 \right>$ as
\begin{equation}
\Delta^2(k) = \frac{k^3}{2\pi^2} P(k) .
\label{DtoPrel}
\end{equation}

\subsection{Correcting for redshift distortions and bias}
\label{zcorr_section}

Kaiser (1987) gave an analytical form for the enhancement of the power
spectrum due to the collapse of waves in the linear regime. For a
single wave, the relation between real space $\Delta_{\rm r}^2$ and redshift
space $\Delta_{\rm z}^2$ is
\begin{equation}
\Delta_{\rm z}^2(k)=\Delta_{\rm r}^2(k)\left[ 1+\mu ^2\frac
fb\right]^2 ,
\label{Kaiser1}
\end{equation}
where $f$ is the linear growth rate of velocity
\begin{equation}
f(\Omega,\Lambda)\equiv\frac{{\mathrm{d\ln}}\,\delta}{{\mathrm{d\ln}}\,a}%
\approx\Omega^{0.6},
\end{equation}
$\mu$ is the direction cosine between the wave vector and the line of sight,
and a scale-independent bias $b$ is included. Averaged uniformly over
all $\mu$, this relation becomes
\begin{equation}
\Delta_{\rm z}^2(k)=\Delta_{\rm r}^2(k)\left[ 1+\frac 23\frac fb+\frac
15\frac{f^2}{b^2}\right] .
\label{Kaiser2}
\end{equation}

This formula describes the modification of the power spectrum on
linear scales.  On nonlinear scales, large random peculiar velocities
within collapsed objects
cause an apparent elongation in a direction along the line of sight
(`fingers-of-God'), which decreases power on cluster-sized scales
and smaller. Assuming that the velocity distribution is Gaussian, this
modifies the power spectrum as
\begin{equation}
\Delta_{\rm z}^2(k)=\Delta_{\rm r}^2(k)\exp \left[ -\frac 12\left( \frac{\mu
k\sigma }{H_0}\right) ^2\right]
\label{finger1}
\end{equation}
where $\sigma$ is the 1D rms velocity dispersion (Peacock 1992). When
averaged over all angles, this gives
\begin{equation}
\Delta_{\rm z}^2(k)=\Delta_{\rm r}^2(k)\left[
\frac{\sqrt{\pi}}2\frac{{\rm erf} (k\sigma/H_0)} {k\sigma/H_0}\right] .
\label{finger2}
\end{equation}

PD94 combined Eq.\ (\ref{Kaiser2}) and (\ref{finger2}) along with the
galaxy bias into a single formula that is applicable on all scales and
also allows for cross-correlation between power spectra from two
different catalogues. This formula is
\begin{equation}
\Delta_{\rm r}^2(k)=\Delta_{\rm z}^2(k)\left[ b_1b_2G(y,\alpha_1,\alpha_2)%
\right] ^{-1}
\label{rbcorr}
\end{equation}
where
\begin{equation}
y\equiv \frac{k}{100}\sqrt{(\sigma_1^2+\sigma_2^2)/2} ,
\label{ydef}
\end{equation}
\begin{equation}
\alpha \equiv f(\Omega )/b ,
\label{alphadef}
\end{equation}
\begin{equation}
\ifref \else \begin{array}{lr}\lefteqn{ \fi        
G(y,\alpha_1,\alpha_2)=%
\ifref \else }&\\& \fi                             
\frac{\sqrt{\pi }}8\frac{{\rm erf}(y)}{y^5}
\left[ 3\alpha_1\alpha_2+2(\alpha_1+\alpha_2)y^2+4y^4\right]
\ifref \else \\& \fi                               
-\frac{\exp (-y^2)}{4y^4}\left[ \alpha_1\alpha_2(3+2y^2)+2(\alpha_1+\alpha_2)%
      y^2\right],
\label{Gdef}
\ifref \else \end{array} \fi                       
\end{equation}
and subscripts 1 and 2 refer to quantities for two different
catalogues, in the case of a cross-correlation. Eq.\ (\ref{ydef}) is
for $k$ in units of \hmpcinv\ and $b$ is the bias.  Eq.\
(\ref{rbcorr}) is to be applied point by point to the raw
observational power spectrum in redshift space.  In the case of the
APM spectrum, $G$ must be taken as unity since there are no
redshift-space distortions.

\begin{figure}
\ifref\else
\resizebox{\figwidth}{!}{\includegraphics{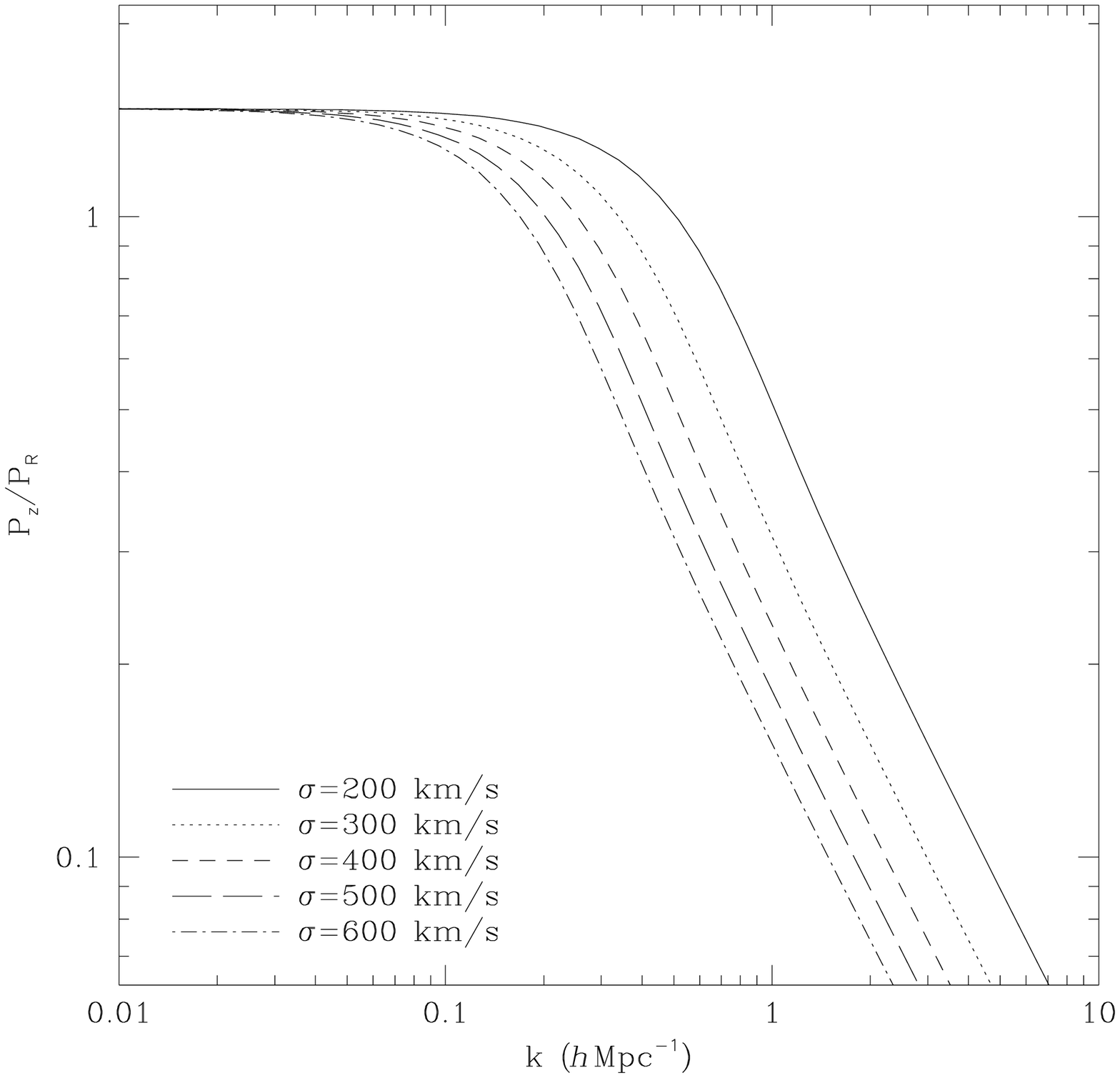}}
\fi
\caption{The redshift correction is plotted as a function of $k$ and
of the 1D velocity dispersion, $\sigma$, for the T\lcdm$_{0.4}$ model
with $b=1$.  On large scales the overall correction is that of Kaiser (1987)
while on small scales corrections are due to the `finger-of-God'
effect, which is sensitive to $\sigma$.  Notice that in the region
$k\protect\ga 0.1$ \hmpcinv, a slightly higher or lower value could alter the
correction noticeably.}
\label{rcorr_fig}
\end{figure}

Because of Eq.\ (\ref{ydef}) and (\ref{Gdef}), the redshift correction
in Eq.\ (\ref{rbcorr}) is sensitive to the velocity dispersion.  PD94
make the approximation that $\sigma=300$ \kms\ for each of the
catalogs in the data set.  We note that a slightly different choice of
$\sigma$ could significantly affect the redshift correction at $k\ga
0.1$ \hmpcinv.  In Figure~\ref{rcorr_fig}
we show how Eq.\
(\ref{rbcorr}) appears when several different values of the velocity
dispersion are chosen for the T\lcdm$_{0.4}$ model.
For example, the correction at $k=1$ \hmpcinv is about 40 per cent
greater for $\sigma = 400$ \kms\ than for $\sigma = 300$ \kms.

\begin{figure}
\ifref\else
\resizebox{\figwidth}{!}{\includegraphics{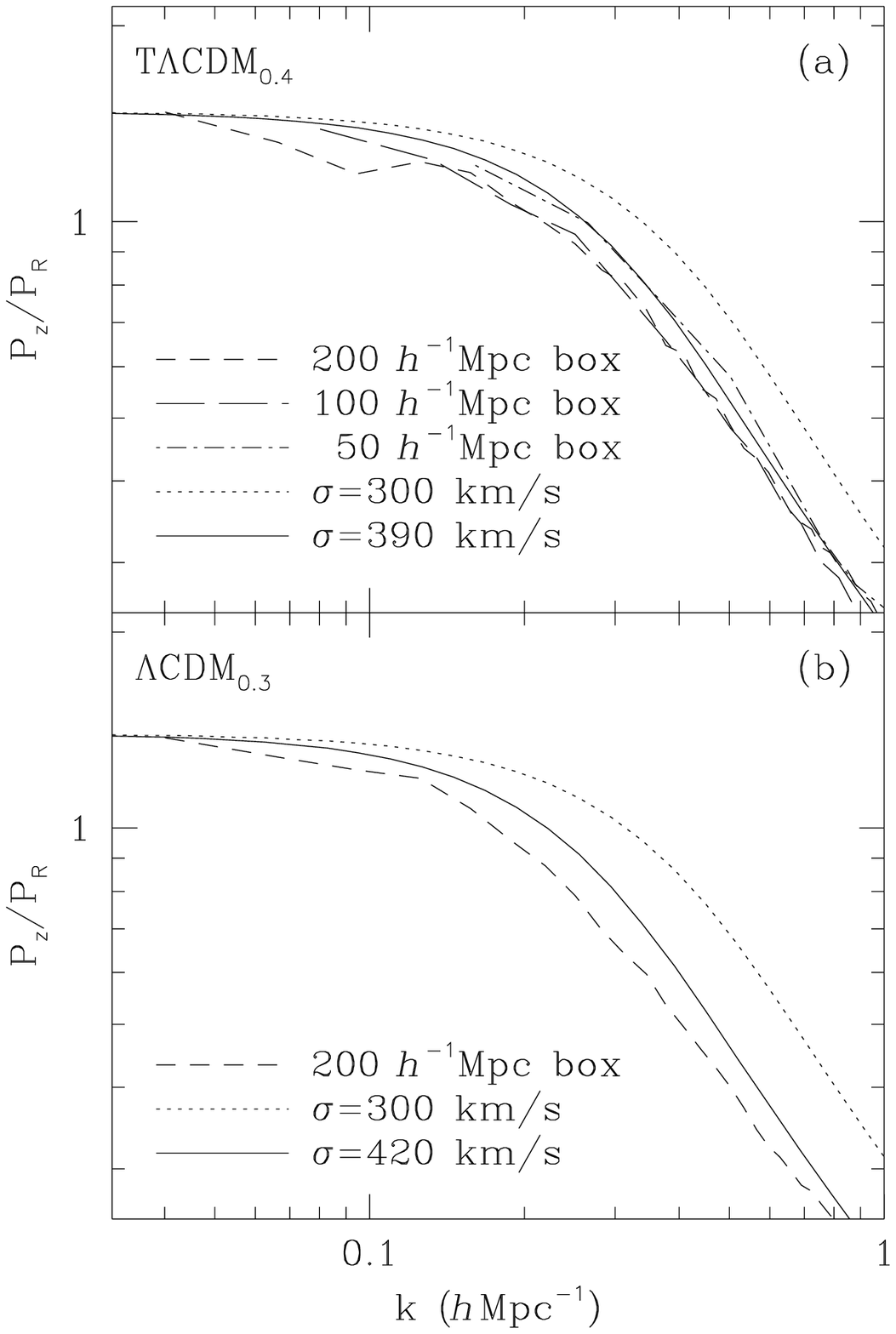}}
\fi
\caption {The redshift correction vs.\ $k$ as verified with
simulations.  Two models were tested, (a) T\lcdm$_{0.4}$ and (b)
\lcdm$_{0.3}$, both with $b=1$.  The results for different box sizes
(broken lines) are plotted along with the analytical predictions of
Eq.\ (\protect\ref{rbcorr}) (solid and dotted lines) for two
different values of $\sigma$.  It can plainly be seen that the
corrections based on $\sigma$ taken from the simulations give much
better fits than the straight $\sigma=300$ \kms\ of PD94.}
\label{simrcorr_fig}
\end{figure}

We are able to test the accuracy of the redshift corrections using our
simulations.  We placed dark matter particles from the T\lcdm$_{0.4}$
model at the $z=0$ moment into redshift space according to Hubble's
law and the peculiar motions of the particles.  Then we calculated the
power spectrum from two different vantage points within the box and
averaged together all of these power spectra from each of the
realizations of a particular box size.  By taking the ratio of the
redshift space power spectrum to the real space power spectrum, we
produced a redshift correction for the model which can be directly
compared with the prediction of Eq.\ (\ref{rbcorr}), which is
presented in Figure~\ref{simrcorr_fig}.
Two different theoretical
predictions are shown, the PD94 prediction with $\sigma = 300$ \kms\ 
and a prediction curve using the average velocity dispersion from
simulations of the largest boxes for each model.  Clearly, the
simulated corrections conform better to the prediction using the
actual velocity dispersion than the PD94's assumed value of $\sigma =
300$ \kms.  We conclude that this is an important effect which {\it
must} be included in this type of analysis, and we will therefore
apply redshift corrections based on the 1D velocity dispersions from
simulations throughout this paper.

In order to arrive at 1D velocity dispersions that are appropriate for
the models that we test in this paper, we examined all models that
have 300 \hmpc\ boxes.  We calculated the rms velocity
dispersion of all particles in the boxes and computed the redshift
correction to the power spectrum as describe above.  We found the
following 1D velocities to be representative of the three families of
models considered: 610 \kms\ for SCDM, 430 \kms\ for \lcdm, and 540 \kms\ 
for CHDM.  These numbers are used for $\sigma$ throughout the rest of
the paper.

\subsection{Correcting for nonlinear growth}

By extending the method of HLKM
from the correlation function to the power spectrum, PD94 arrived at a
fitting formula that would relate the linear to the nonlinear
spectrum. This procedure has two parts: adjusting the amplitude of the
power spectrum, and transforming between the linear and nonlinear
scales. Originally it was thought that the linear--nonlinear mapping
would be nearly model-independent, with the density parameters being
the only information about the model to enter into the equations (cf.
PD94). However, later investigation by Jain, Mo, \& White (1995)
showed that there was a significant dependence on the slope of the
linear power spectrum as well.  We chose to adopt the new method of
PD96 with small modifications for performing the linear-nonlinear
mapping.

We assume that the linear $\Delta^2_{\rm L}(k)$ is related to the
nonlinear $\Delta^2_{\rm NL}(k)$ through
\begin{equation}
\Delta^2_{\rm NL}(k_{\rm NL})=f_{\rm NL}\left(\Delta^2_{\rm L}(k_{\rm L})\right)
\label{PD96usefit}
\end{equation}
where $f_{\rm NL}$ is a fitting function of the form
\begin{equation}
f_{\rm NL}(x)=x\left[\frac{1+B\beta x+\left[ Ax\right] ^{\alpha \beta }}
{1+\left( \left[ Ax\right] ^\alpha g^3(\Omega )/\left[ Vx^{1/2}\right]
\right) ^\beta }\right] ^{\frac 1\beta } .
\label{PD96fit}
\end{equation}
This formula has an explicit dependence on $\Omega$ through the growth
suppression factor $g(\Omega,\Lambda)$, given in Eq.\ (\ref{gOmdef}), and
five free parameters.  The best-fitting parameters were determined by
PD96 to be
\begin{equation}
\label{paramstart}
A=0.482(1+n/3)^{-0.947} ,
\label{paramA}
\end{equation}
\begin{equation}
B=0.226(1+n/3)^{-1.778} ,
\label{paramB}
\end{equation}
\begin{equation}
\alpha =3.310(1+n/3)^{-0.244} ,
\label{paramalpha}
\end{equation}
\begin{equation}
\beta =0.862(1+n/3)^{-0.287} ,
\label{parambeta}
\end{equation}
\begin{equation}
V=11.55(1+n/3)^{-0.423} ,
\label{paramend}
\end{equation}
which are all dependent on the local slope of the linear power
spectrum,
\begin{equation}
n\equiv \frac{d\ln P}{d\ln k} .
\label{slopedef}
\end{equation}

In addition to a change of amplitude there must also be a change of
scale.  As linear perturbations become nonlinear, objects with density
contrast $1+\delta$ have collapsed radially by a factor of $(1+\delta
)^{1/3}$. This suggests that the scale $k_{\rm NL}$ is related to the
scale $k_{\rm L}$ by the equation
\begin{equation}
k_{\rm NL}=\left[ 1+\Delta_{\rm NL}^2(k_{\rm L})\right]^{1/3}k_{\rm L} .
\label{nlscaledef}
\end{equation}
Taken together, Eqns.\ (\ref{PD96usefit}) and (\ref{nlscaledef}) and
their attending equations comprise the nonlinearization process, which
predicts the observed nonlinear power spectrum in redshift space from
a given linear spectrum in real space.

A more useful applications of these methods would be the inverse
process -- linearization.  Eq.\ (\ref{rbcorr}) can simply be solved
for $\Delta_{\rm r}^2(k)$ and the fitting formula in Eq.\ (\ref{PD96fit})
can be inverted numerically.  The change of scale expressed in Eq.\
(\ref{nlscaledef}) can be stated equivalently as
\begin{equation}
k_{\rm L}=\left[1+\Delta_{\rm NL}^2(k_{\rm NL})\right]^{-1/3}k_{\rm NL} ,
\label{lscaledef}
\end{equation}
completing the tool set for use in manipulating the power spectrum.

This method works well for scale-free power law spectra with a
constant slope $n$ of the linear $P(k)$, but not for scale-dependent
power spectra.  The curving shape of CDM-type power spectra could not
be reconstructed with much success.  Jain et al. (1995) suggested that
CDM-type spectra could be treated by choosing a constant `effective'
slope, $n_{\rm eff}$, to take the place of $n$ in Eqns.\ 
(\ref{paramstart}--\ref{paramend}).
The effective slope is defined as the tangent slope of the linear
$P(k)$ at the nonlinear scale (where the mass fluctuation is unity).
A linearization using $n_{\rm eff}$ can only approximate the true shape of
the linear spectrum since the method is based on deriving the linear
power from the nonlinear power on a smaller scale.  Because the slope
of the CDM-type spectrum is always changing, the effective index
should continually change with scale, as was recognized by PD96.  We
choose $n$ from the linear power spectrum on the appropriate linear
scale $k_{\rm L}$ by numerically differentiating $P(k_{\rm L})$, which works
reasonably well.

\section{Application of the method to simulations}
\label{appsims}

In order to thoroughly test this method, we applied it to our database
of simulations or ran new simulations suitable for this purpose.  For
each realization listed in Table~\ref{simsum_table}, we calculated the
nonlinear power spectrum at the moment $z=0$ and then averaged them
together for each box size.  For the T\lcdm$_{0.4}$ model, which has
multiple box sizes, we constructed a complete nonlinear power spectrum
by combining the spectra starting with the power from the largest box
size and adding power from successively smaller boxes multiplied by a
small factor if needed (to account for power from missing long waves).

\begin{figure*}
\ifref\else
\resizebox{\textwidth}{!}{\includegraphics{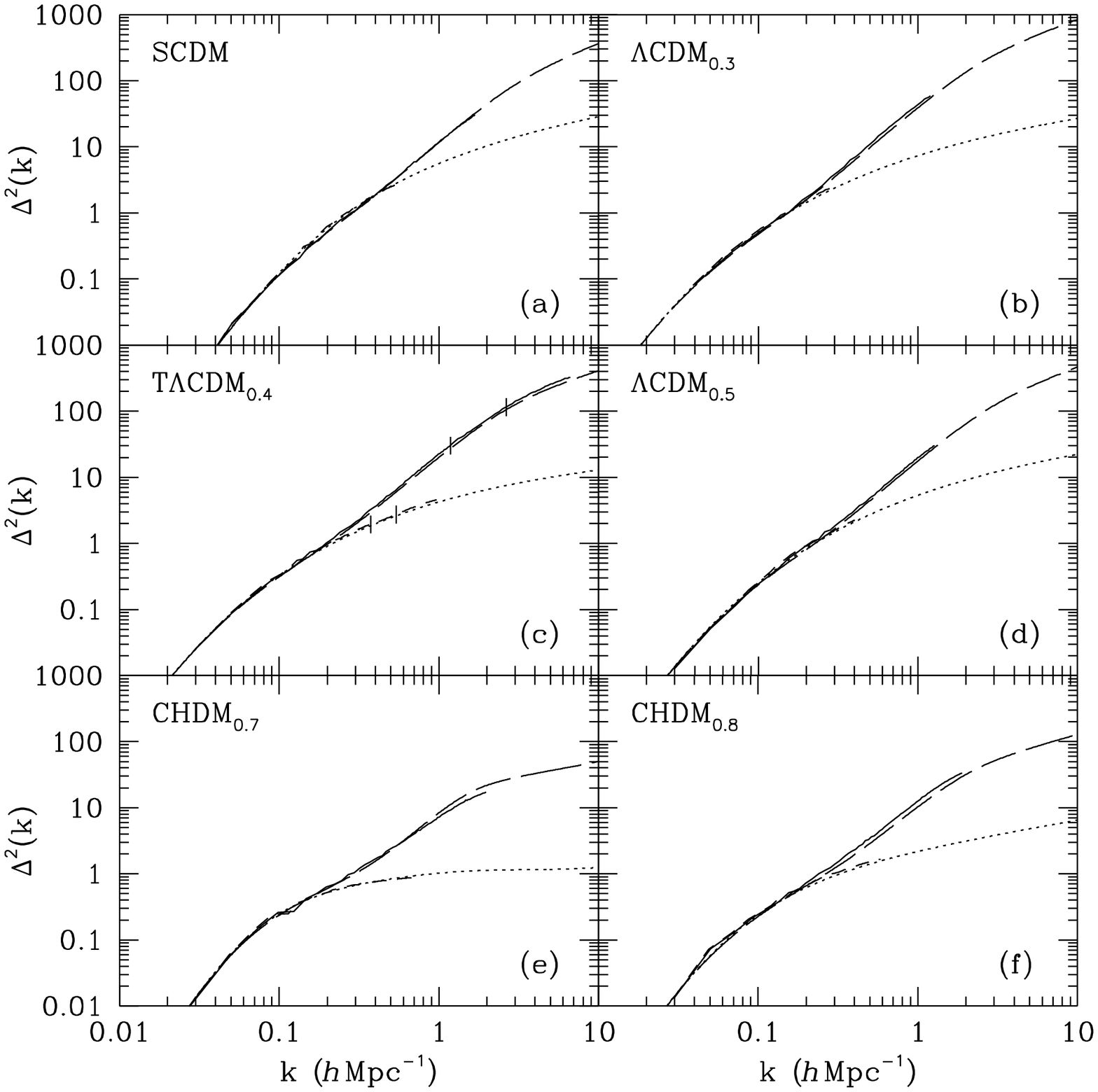}}
\fi
\caption {This figure plots the linear power spectrum for each model
(dotted line) and the nonlinear power spectrum from simulations (solid
line) against the linearized power spectrum (short dashed line) and
the nonlinearized power spectrum (long dashed line).  The linearized
and nonlinearized lines are from the application of the method
described in \S\protect\ref{method}, with a modification of the local
slope of the linear power spectrum
$n_{\mathrm{L}}(k_{\mathrm{L}})\longrightarrow n_{\mathrm{L}}(k_{\mathrm{L}}/2)$
as used in PD96.
Because the T\lcdm$_{0.4}$ model has simulations with high resolution
that sample smaller waves, it is plotted to higher values of $k$.  The
short horizontal lines denote
boundaries between the 300, 100, and 50 \hmpc\ 
samples.  At high values of $k$ resolution effects in the
simulations bring the power spectrum lower than what it should be.  We
chose to stop plotting curves once $\Delta^2(k)$ turned over.}
\label{linsim_fig}
\end{figure*}

With this information we could test the procedure in either of two
ways: linearizing the nonlinear power spectrum and then comparing it
to the linear power spectrum of the model, or nonlinearizing the
generated linear power spectrum and then comparing it to the actual
nonlinear spectrum from simulations.  In fact, we did both so that we
could demonstrate the consistency of both approaches. We tested the
linear--nonlinear corrections of \S\ref{method}.2 on the final
nonlinear and linear power spectra of the models.  The results are
presented in Figure~\ref{linsim_fig}
for all of the models that we
used with the exception of \lcdm$_{0.4}$ (since it is very similar to
T\lcdm$_{0.4}$).  These plots incorporate the shifted mapping
$n_{\mathrm{L}}(k_{\mathrm{L}})\longrightarrow n_{\mathrm{L}}(k_{\mathrm{L}}/2)$
from PD96.  All models appear to fit quite well over the entire
range in scales that we have been able to faithfully simulate.

\begin{figure*}
\ifref\else
\resizebox{\textwidth}{!}{\includegraphics{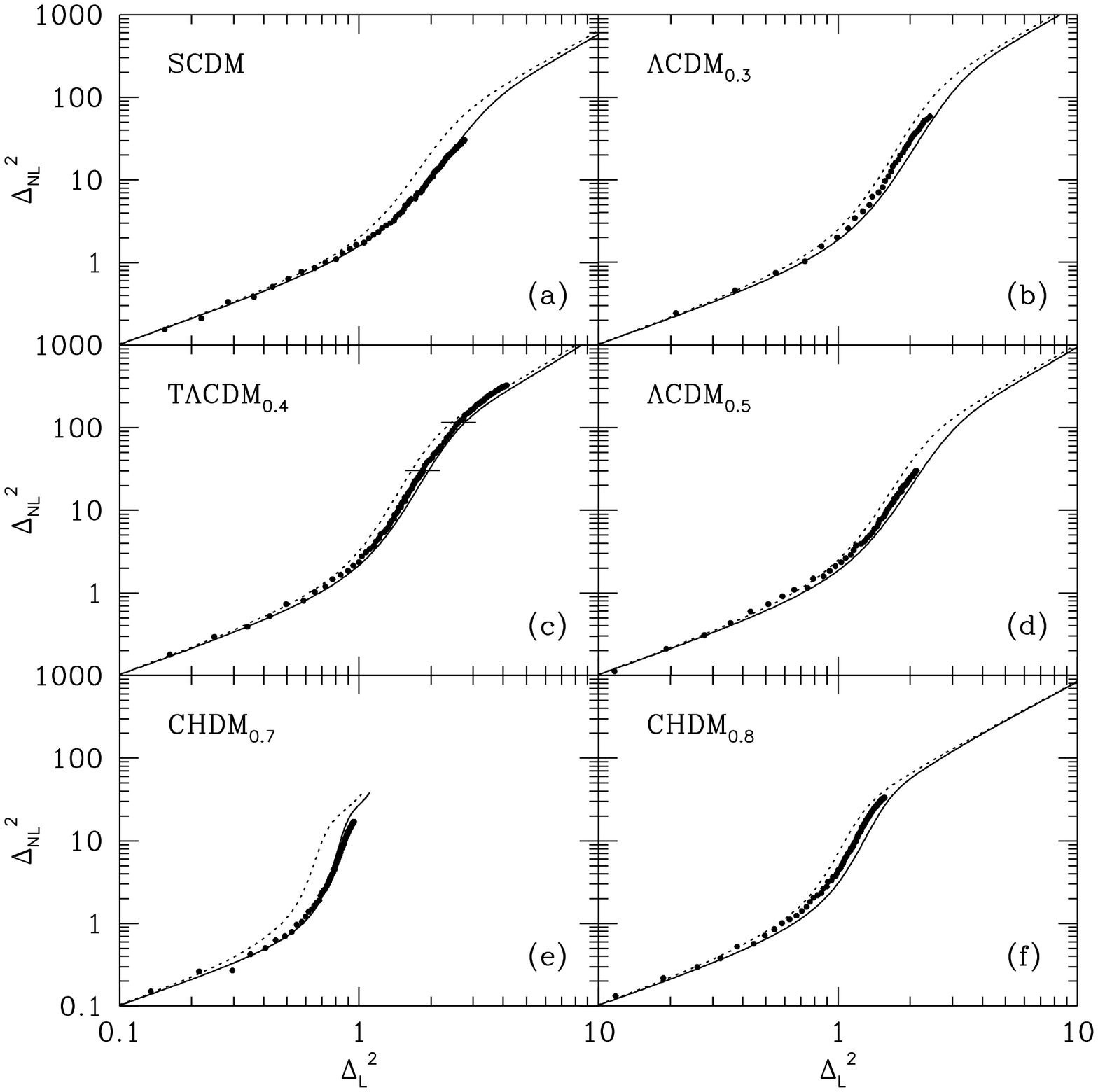}}
\fi
\caption {The effects of using shifted or unshifted $n$ on the
linear--nonlinear mapping.  The solid line represents the mapping in
Eq.\ (\protect\ref{PD96fit}) performed without a shift while the
dotted line represents the mapping with a shift of a factor of 2, as
in PD96.  The filled circles show the linear--nonlinear relation of
the real-space power spectrum from simulations of the models.}
\label{n_effects_fig}
\end{figure*}

Figure~\ref{n_effects_fig}
presents the relation between
$\Delta_{\rm NL}^2$ and $\Delta_{\rm L}^2$.  This format removes $k$ dependence
by plotting the linear-nonlinear mapping only.  This allows a more
careful examination of the goodness of fit than the previous Figure.
In addition to plotting the shifted linear--nonlinear mapping of PD96,
the mapping with no shift in $n_{\rm L}(k_{\rm L})$ is plotted for comparison.
We find that most models favor the shifted mapping except for \lcdm$_{0.4}$
model which clearly favored no shift, for which we have no explanation.

\section{Application of the method to observations}
\label{appdata}

Now that a reliable linearization method has been designed and tested,
it can be applied to real data in an attempt to constrain the models
that best represent the Universe.  As opposed to linearizing the
observational data based on an arbitrary model and then comparing the
linear power of a model in question with the linearized spectrum
(PD94, Ma 1996), a particular model was applied to the observational
data and the linearization was performed according to the specific
theoretical predictions of that model.  The most correct model should
be identifiable by self-consistency: an exceptionally good fit of the
linearized observations to the theoretical linear spectrum.

\begin{figure*}
\ifref\else
\resizebox{\textwidth}{!}{\includegraphics{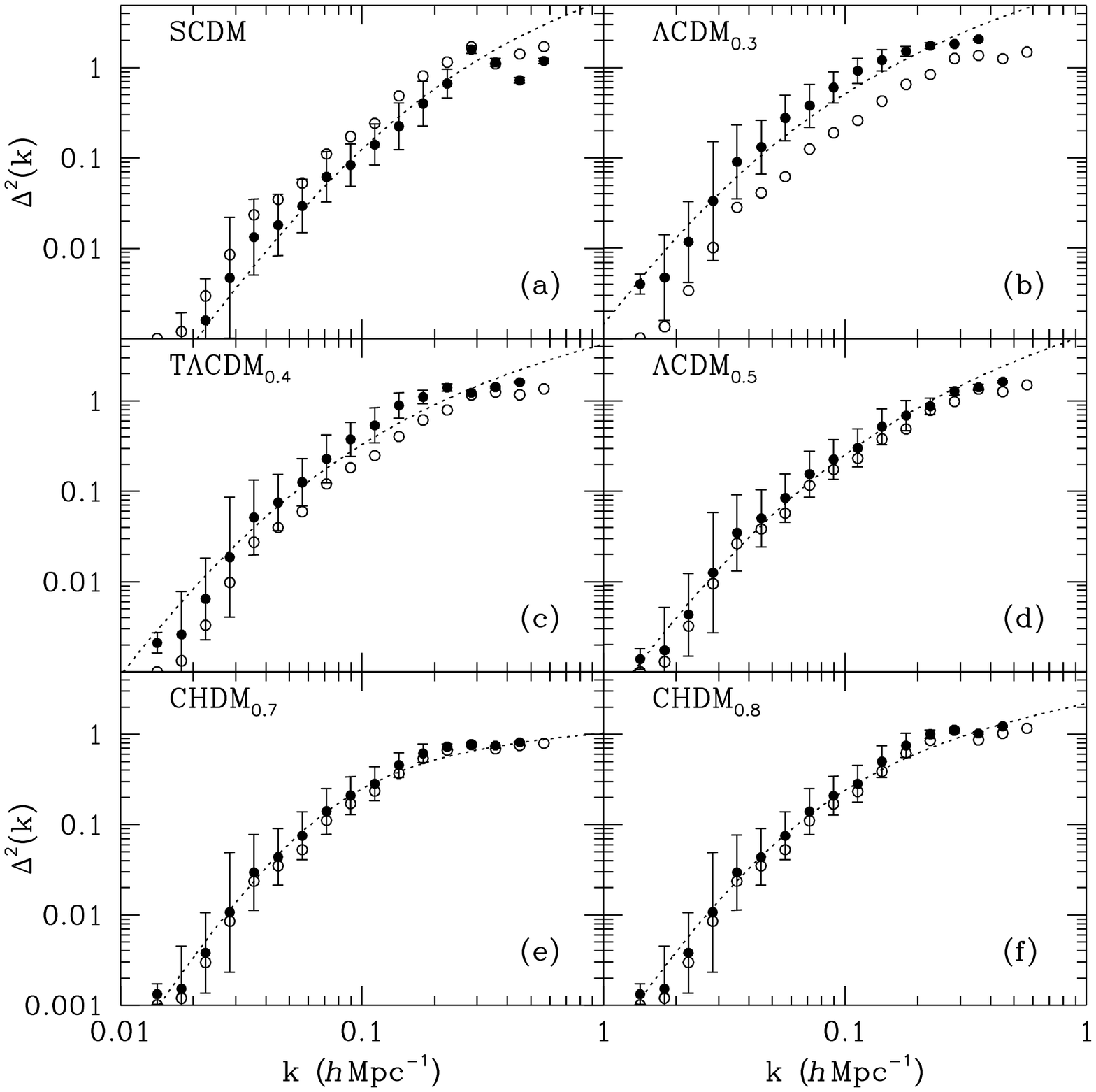}}
\fi
\caption {The application of the linearization technique to the
observational data set. The linear power spectrum of the imposed model
(dotted line) and two different cases of bias parameters are plotted
(dots with error bars). The open dots correspond to the bias level of
PD94 (Eq.\ (\protect\ref{biasratio}) in this paper) and the filled
dots are for the bias level adjusted by the factor given in Table~%
\protect\ref{fitsum_table}, as determined by minimizing $\chi^2$.
Error bars represent internal errors of each point in the bin added in
quadrature.  See the text for a more detailed explanation.}
\label{linobs_fig}
\end{figure*}

We adopted the same compilation of observational power spectra and
error estimates as in PD94.  For each data set we applied the
linearization procedure for each cosmological model to correct for
redshift distortions (except in the case of APM data), bias, and
nonlinear effects.  Once all data sets were linearized, the collective
sample of power spectrum data points were averaged in bins
$0.1\,\log_{10} k$ wide.  The bin widths and centers were chosen to be
nearly identical to those of PD94 in their Table~1.  The resulting
plots appear in Figure~\ref{linobs_fig} and are discussed below.

Our approach in treating errors is different from PD94.  They simply
accepted errors based on counting statistics in bins, thus neglecting
internal errors of the data sets.  We attempted to keep track of
individual error bars throughout the linearization process, assuming
that errors of points in each data set are independent.  This is not
actually true, but it is difficult to estimate how the data points
correlate.  The error estimates were made in the following way. When
performing redshift and bias corrections, the data points were simply
scaled up or down by a factor (Eq.\ (\ref{rbcorr})), so the error
estimates were simply scaled by the same factor.  For the case of
nonlinear corrections, we performed the correction on both $\Delta^2$
and the upper limit $\Delta^2+\delta$ and used the difference between
these two corrected points as the error estimate. Technically, this
gives an error at a different $k$ than the data point in question,
which has the effect of reducing the size of the error.  However,
since the original errors are uncertain to begin with, we expect that
our final estimates are at least a fair representation of the errors.
Correlations in the observational data points will lead to the plotted
errors being an underestimate of the true errors.

While we assumed the same relative biases as PD94 (Eq.\
(\ref{biasratio})), the biases were all scaled by the same constant so that
the linearized spectrum obtained from observational data could be best
fit to the linear dark matter spectrum according to a $\chi^2$ test.
Each bias in Eq.\ (\ref{biasratio}) was multiplied by this factor,
reflected in the value of $b_{\rm I}$, before being applied to the
observational data through Eq.\ (\ref{rbcorr}).  Table~\ref{fitsum_table}
\begin{table}
\caption {Summary of best-fit biases}
\label{fitsum_table}
\begin{center}
\begin{tabular}{lcccc}
Model & $b_{\rm I}^a$ & $1.3 b_{\rm I}^b$ & $b_{\rm O}^c$ & $P$ ratio$^d$ \\
\hline
SCDM & 1.42 & 1.85 & $1.35 \pm 0.24$ & 0.98 \\
\\
$\Lambda $CDM$_{0.3}$ & 0.50 & 0.64 & $0.69 \pm 0.12$ & 1.41 \\
$\Lambda $CDM$_{0.4}$ & 0.58 & 0.75 & $0.81 \pm 0.14$ & 1.34 \\
T$\Lambda $CDM$_{0.4}$ & 0.69 & 0.90 & $0.81 \pm 0.14$ & 1.33 \\
$\Lambda $CDM$_{0.5}$ & 0.85 & 1.11 & $0.92 \pm 0.16$ & 1.09 \\
\\
CHDM$_{0.7}$ & 0.87 & 1.13 & $-$ & 0.99 \\
CHDM$_{0.8}$ & 0.87 & 1.13 & $-$ & 1.04 \\
\hline
\end{tabular}
\end{center}

$^a$ Factor by which the PD94 bias must be multiplyed to achieve best fit.

$^b$ Bias of blue galaxies as indicated by the best fit $b_{\rm I}$.

$^c$ Bias of blue galaxies from \potent\ analysis 
Hudson \etal\ (1995).

$^d$ Average of $P_{\rm linear}/P_{\rm linearized}$ at
$k=$0.071, 0.090, and 0.113 \hmpcinv.
\end{table}
gives the best $b_{\rm I}$ for each model as well as an
indicator meant to quantify the goodness of fit of the linearizations.
The indicator (called the `$P$ ratio') is the average of the ratio
of linear power to linearized power at three points near the middle of
the sample, $k=$0.071, 0.090, and 0.113 \hmpcinv, chosen to be near
the maximum curvature of the power spectra and have relatively small
error bars.  Since decreasing $b_{\rm I}$ moves the spectrum up and
increases its curvature, this indicator should be a reliable estimate
of how close the curvature of the linearized spectrum is to matching
that of the linear power spectrum for the model.  In the case where
the curvature is very different, however, the $P$ ratio tends to be
low (since the two best-fitting curves intersect at the midpoint)
giving a false indication of a good fit.  In our sample of models, the
$P$ ratio for the SCDM model is the only one to suffer from this
problem because the SCDM power spectrum simply has the wrong shape.

Note that the last few (i.e., highest $k$) binned points are always
less reliable than the error bars indicate.  They are either averages
of only the APM and CfA results or of APM alone and have few points
per bin. Quite often (e.g., Liddle \etal\ 1996a)
the highest-$k$ points are ignored because the earlier linearization
algorithms were not accurate for these points.  However, we think that
these points should be included in the analysis as long as there are
enough data points in the bins. They have the smallest statistical
error bars and now cannot be excluded solely on the basis of failure
of the linearization procedure, since our method demonstratably works
up to this range in $k$.

We find that the \lcdm\ family of models have the wrong shape when the
best fitting bias $b_{\rm I}$ is determined.  All of these models have too
much curvature and fit poorly at the high-$k$ tail of points, even
though most of the points are within the nominal error bars.  As $b_{\rm I}$
is lowered below unity, the curves of the reconstructed linear power
spectra rise in amplitude and increase in curvature, which is what is
responsible for the relatively poor fits.  The reconstructed CHDM
models both seem to do an exceptionally good job of fitting their
respective power spectra.  The excellent match of the curvature and
the small adjustment to the bias make CHDM the most successful class
of candidate models in our sample.

\begin{figure}
\ifref\else
\resizebox{\figwidth}{!}{\includegraphics{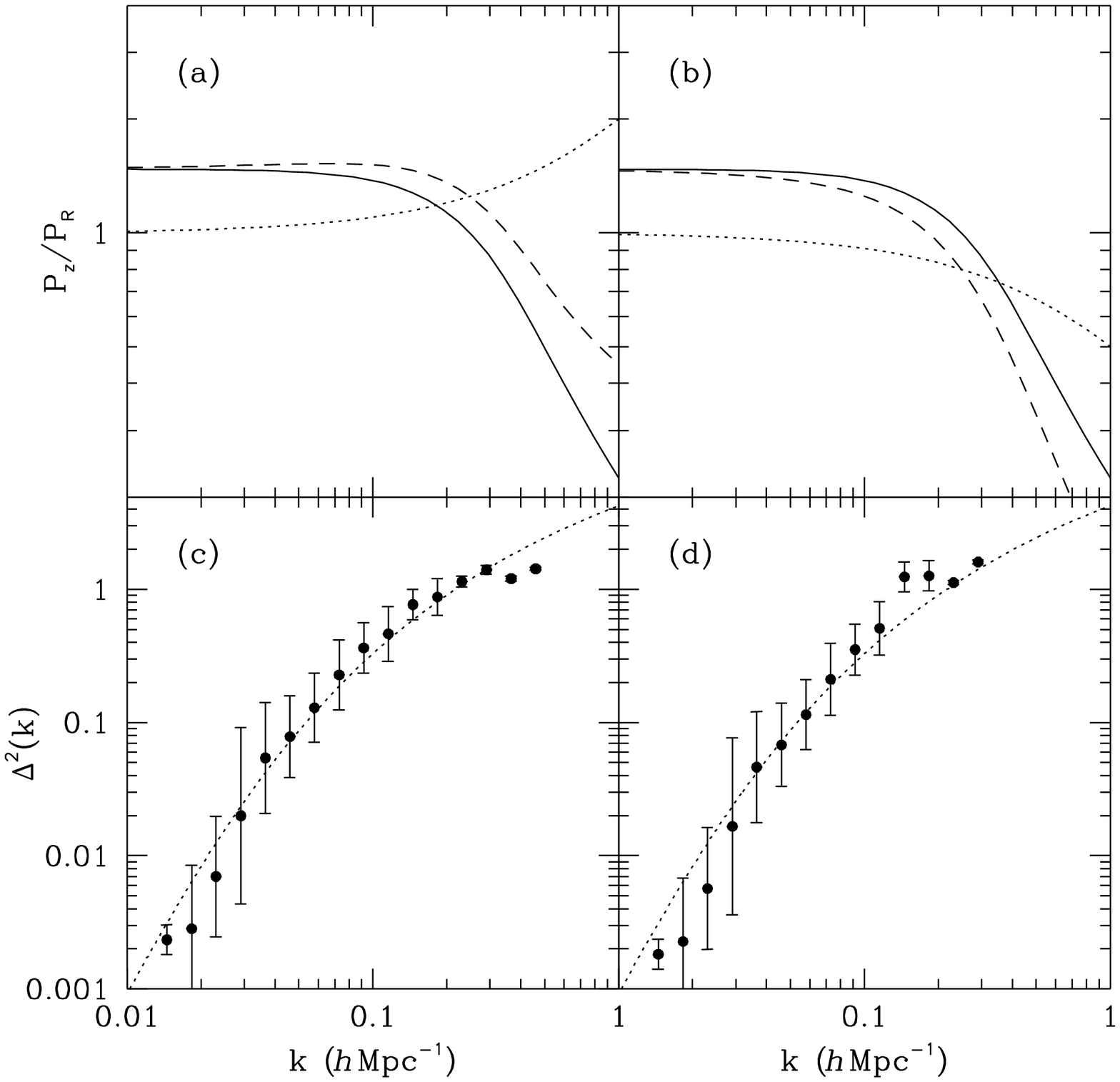}}
\fi
\caption {Test forms of scale-dependent bias and their effects on the
recovered linear power spectrum for T\protect\lcdm$_{0.4}$.  The solid
line is the redshift correction of Eq.\ (\protect\ref{rbcorr}), the
dotted line is the scale-dependent bias function and the dashed line
is the application of the function to the redshift correction.  (a)
Bias increasing linearly with scale to a factor of 2 at $k=1$.  (b)
Bias decreasing linearly with scale to a factor of 2 at $k=1$.  (c)
The positive bias increases the curvature requiring more antibias.
(d) The antibias decreases the curvature requiring less antibias. }
\label{sdepbias_fig}
\end{figure}

Because the \lcdm\ power spectra did not fit well, we tried examining
the effects of another possible modification process.  The formalism
of \S\ref{method}.1 assumes that the bias does not change with
scale.  This is a useful first approximation, but may prove to be an
oversimplification.  It is quite possible that the bias depends on
scale.  Preliminary results from sophisticated $N$-body plus hydro
simulations of galaxy formation done by Yepes \etal\ (1996)
(which include multi-phase treatment of gas and supernovae feedback)
indicate that bias is slightly rising with $k$ for both CDM and \lcdm\
models.  The semi-analytic merging hierarchy inclusion of hydrodynamic
effect in simulations by Kauffmann, Nusser, \& Steinmetz (1995) gave
similar results regarding scale dependence of bias.  In order to obtain
a rough estimate of the possible
effect we introduce a very simple model for bias which modifies the
redshift-space correction of Eq.\ (\ref{rbcorr}).  Figure~\ref{sdepbias_fig}
shows the form of a bias which increases with scale
in (a) and which decreases with scale in (b).  It is a linear function
of $k$: $b_{\rm I}=1+k_{\rm NL}/(\mbox{1 \hmpcinv})$ which is a factor of two
at $k_{\rm NL}=1$ \hmpcinv.  Panel (c) shows that an extra positive
bias makes the best fit even worse while (d) shows that
anti--bias improves the best fit by raising the high-$k$ tail. Because
such a large anti--bias is difficult to motivate physically, we do not
regard it as an acceptable solution to this problem (cf. KPH96).

\begin{figure}
\ifref\else
\resizebox{\figwidth}{!}{\includegraphics{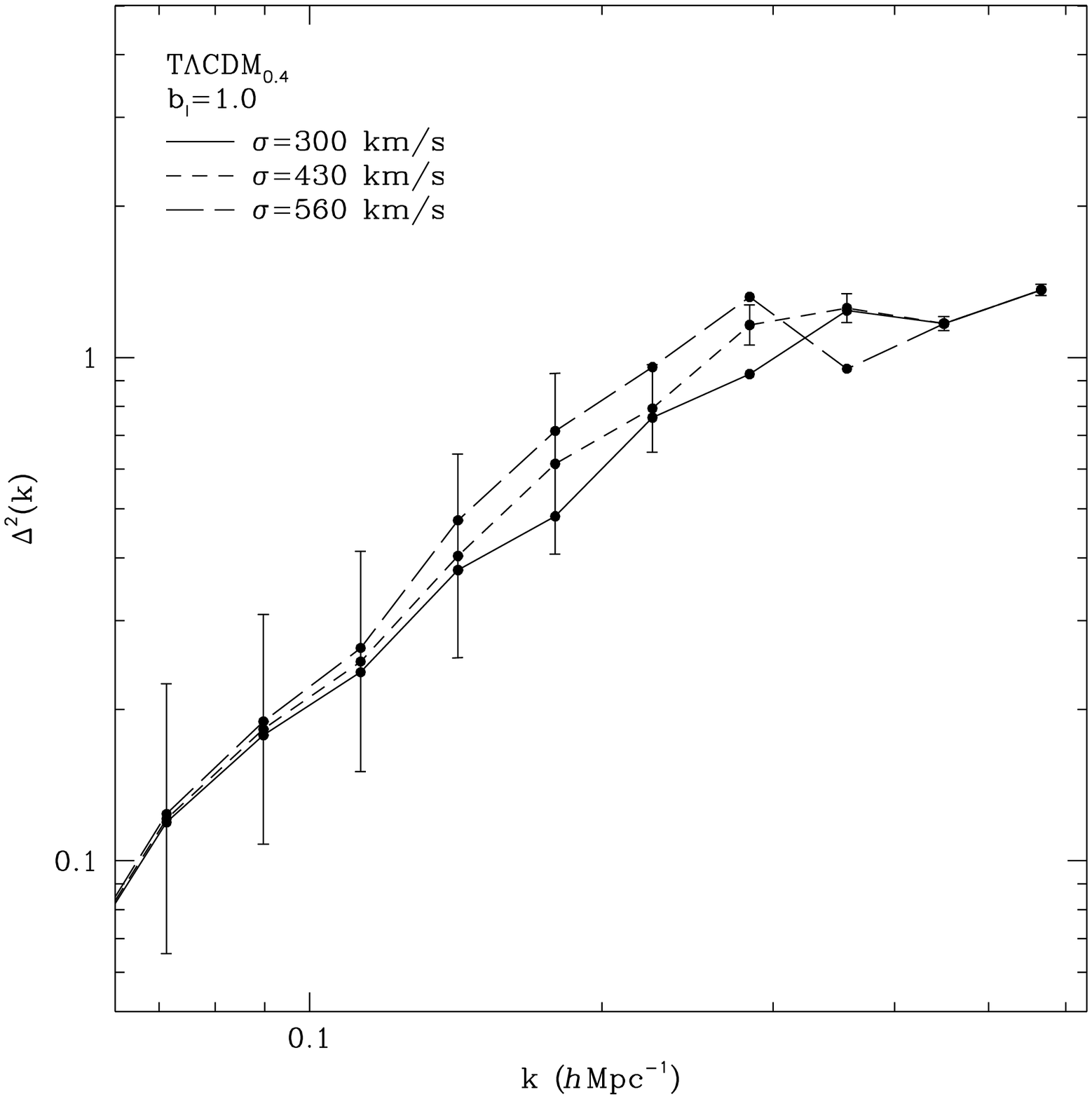}}
\fi
\caption {The effect of different choices of 1D velocity dispersion on
the reconstructed observational T\protect\lcdm$_{0.4}$ power spectrum.
Error bars have been included for the $\sigma=430$ \kms\ case only,
but are representative of the other cases.  Only the high-$k$ points
are shown since the differences are minimal at low $k$ (see Figure~%
\protect\ref{rcorr_fig}).  The last two points are degenerate for all
values of $\sigma$ since they are constructed from APM data only,
which are not affected by the redshift correction.}
\label{zceffects_fig}
\end{figure}

At this point we can demonstrate that the choice of 1D velocity
dispersion has an impact on the form of a reconstructed linear power
spectrum, as was argued in \S\ref{zcorr_section}.  Figure~\ref{zceffects_fig}
shows that the velocity dispersion separates the
different cases at the high-$k$ tail for the unbiased T\lcdm$_{0.4}$
model.  Three realistic values of $\sigma$ were chosen, 300 \kms\ as in
PD94, 430 \kms\ as was used for all \lcdm\ models, and 560 \kms\ 
which is representative of CHDM models.  Since for $k>0.2$ \hmpcinv\ the
curves are separated by an amount greater than the error bars, the
redshift correction can have a noticeable effect on the amplitude of
the reconstructed power spectrum.

\begin{figure}
\ifref\else
\resizebox{\figwidth}{!}{\includegraphics{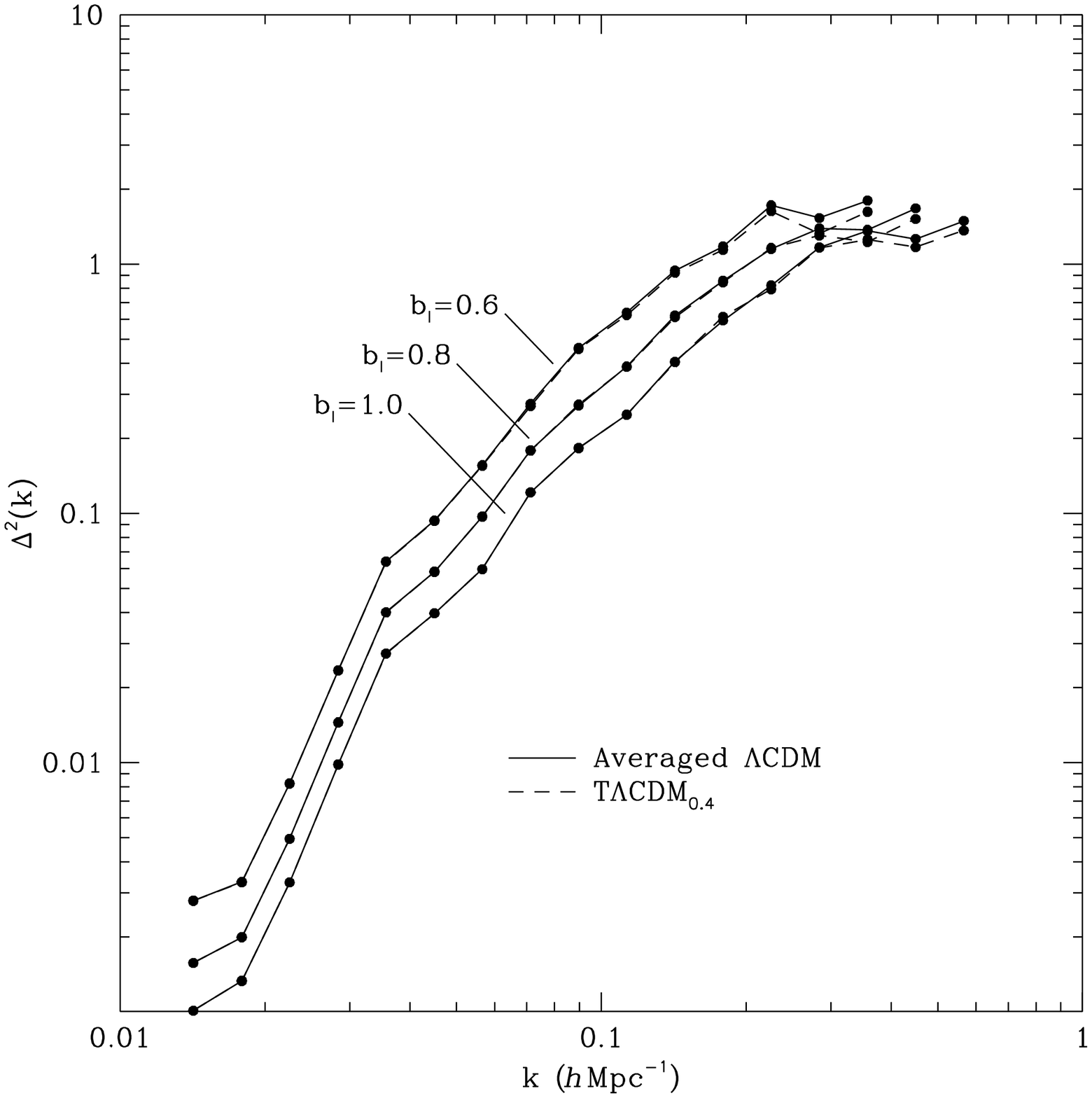}}
\fi
\caption {Observational linear power spectrum reconstructed
under an assumption that the Universe is of a \protect\lcdm\ type. Because the
reconstructed spectra were very similar for \protect\lcdm$_{0.3}$,
\protect\lcdm$_{0.4}$, and \protect\lcdm$_{0.5}$ models, we present
averaged spectra for assumed level of bias.
The spectrum reconstructed assuming the T\protect\lcdm$_{0.4}$ models
differ slightly from the rest and are drawn as the dashed
lines. These data points are given in Table~\protect\ref{lcdmfam_table}.}
\label{lcdmfamily_fig}
\end{figure}

\begin{figure}
\ifref\else
\resizebox{\figwidth}{!}{\includegraphics{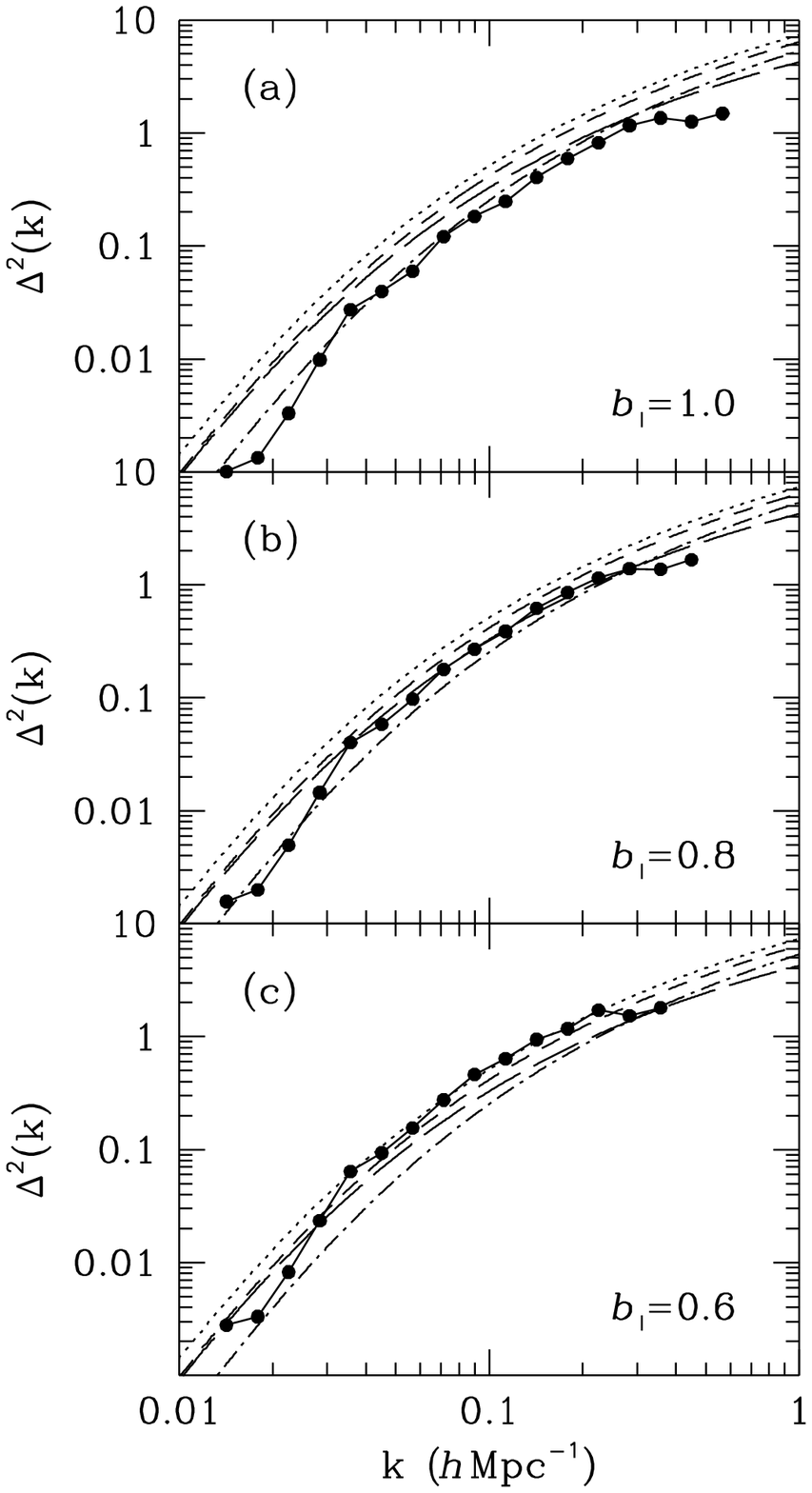}}
\fi
\caption {Comparison of averaged \protect\lcdm$_{0.3}$,
\protect\lcdm$_{0.4}$ and \protect\lcdm$_{0.5}$
simulations with
linearized observational power spectra appropriate to \protect\lcdm\ 
cosmology for different biases. Dotted
lines depict the linear power spectrum of \protect\lcdm$_{0.3}$, short
dashed lines depict \protect\lcdm$_{0.4}$, long dashed lines depict
T\protect\lcdm$_{0.4}$, and dot-dashed lines depict
\protect\lcdm$_{0.5}$.  The biases in the panels are (a) $b_{\rm I}=1.0$,
(b) $b_{\rm I}=0.8$, and (c) $b_{\rm I}=0.6$.}
\label{lcdmcomp_fig}
\end{figure}

An interesting thing that we noticed was that the reconstructed
observational power spectra of our \lcdm\ models were virtually
identical.  The T\lcdm$_{0.4}$ model was the only exception (because
of the tilt), but the difference is small.  Figure~\ref{lcdmfamily_fig},
shows the average of the \lcdm$_{0.3}$,
\lcdm$_{0.4}$ and \lcdm$_{0.5}$ models along with the T\lcdm$_{0.4}$
model for $b_{\rm I}$ values 1.0, 0.8 and 0.6.  Using this to analyze the
fit of the general \lcdm\ spectrum, the averaged reconstructed
\lcdm\ spectrum for the same three $b_{\rm I}$ values overlaid by the linear
power spectra of all of the \lcdm\ models are shown in Figure~%
\ref{lcdmcomp_fig}.
Again, this figure demonstrates that no level of
constant $b_{\rm I}$ can provide an adequate reconstruction of the linear
power spectrum for the \lcdm\ models: the shape of the spectrum is not
optimal.

\begin{figure}
\ifref\else
\resizebox{\figwidth}{!}{\includegraphics{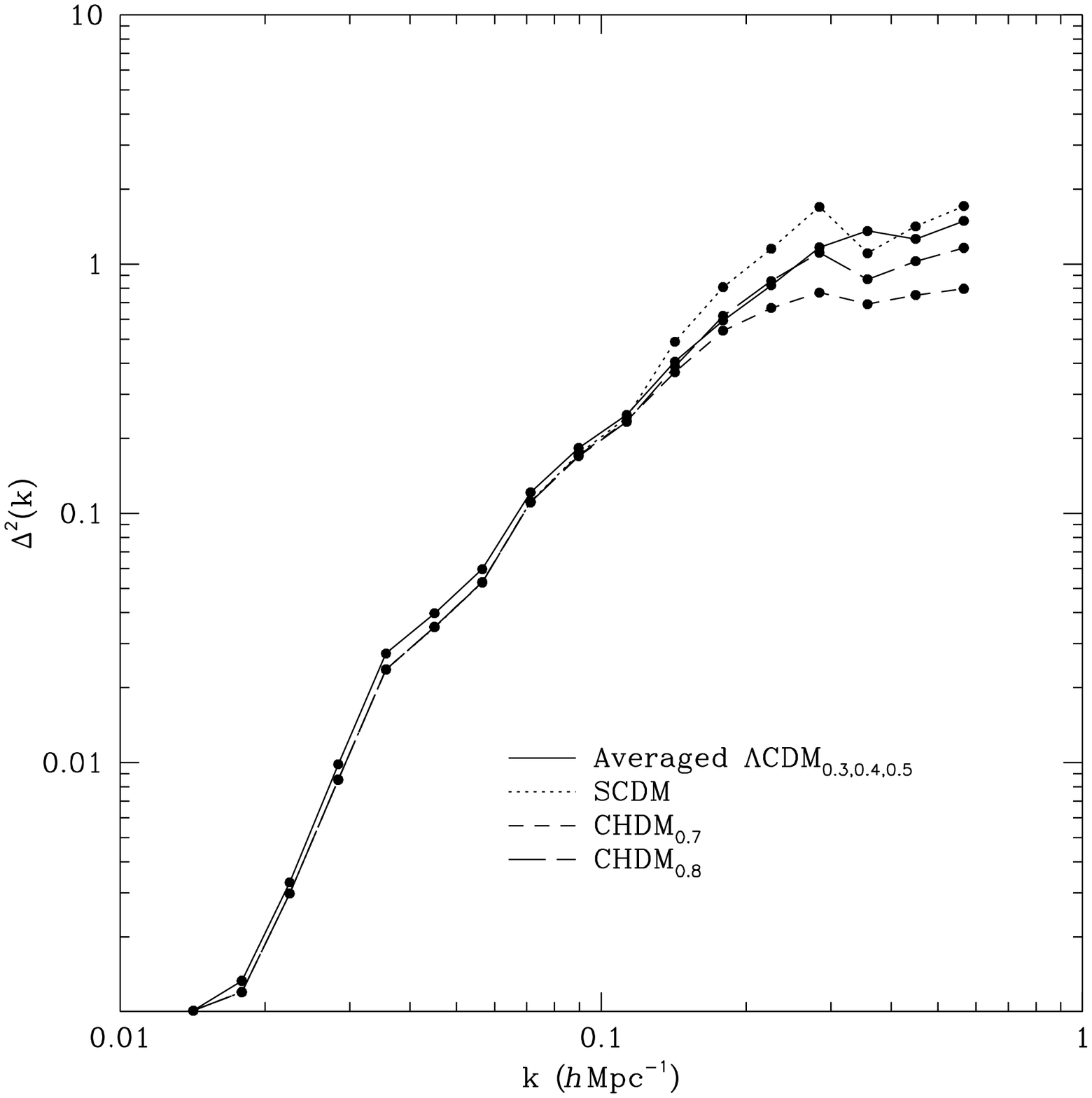}}
\fi
\caption {Comparison of different linearized observational power
spectra with $b_{\rm I} = 1.0$ in each case.}
\label{modelscomp_fig}
\end{figure}

A similar finding is that there is a universal linear
power spectrum reconstructed from the observational data.
Reconstruction for all models that we examined resulted in practically
the same spectrum in the range
$0.01\la k\la 0.1$ \hmpcinv.  Figure~\ref{modelscomp_fig}
shows that all models share an almost identical
shape and amplitude in this region.  This means that any future
progress in distinguishing between these models will need to be
emphasize nonlinear scales $k\ga 0.2$ \hmpcinv. The data shown
in Figure~\ref{modelscomp_fig} have been tabulated in Tables~%
\ref{lcdmfam_table}--\ref{scdmfam_table}.
\begin{table*}
\begin{minipage}{308pt}
\begin{center}
\caption {\protect\lcdm\ family of recovered linear power spectra
\label{lcdmfam_table}
$\Delta^2(k)=\frac{k^3}{2\pi^2}P(k)$}
\begin{tabular*}{308pt}{ccccccc}
& \multicolumn{2}{c}{$b_{\rm I}=0.6$}
& \multicolumn{2}{c}{$b_{\rm I}=0.8$}
& \multicolumn{2}{c}{$b_{\rm I}=1.0$} \\
$k$,(\hmpcinv) & \lcdm & T\lcdm & \lcdm & T\lcdm & \lcdm & T\lcdm \\
\hline
0.0142 & 0.0028 & 0.0028 & 0.0016 & 0.0016 & 0.0010 & 0.0010 \\
0.0179 & 0.0033 & 0.0033 & 0.0020 & 0.0020 & 0.0013 & 0.0013 \\
0.0225 & 0.0082 & 0.0082 & 0.0050 & 0.0049 & 0.0033 & 0.0033 \\
0.0284 & 0.0234 & 0.0234 & 0.0145 & 0.0145 & 0.0098 & 0.0098 \\
0.0357 & 0.0641 & 0.0638 & 0.0400 & 0.0400 & 0.0274 & 0.0273 \\
0.0450 & 0.0935 & 0.0932 & 0.0583 & 0.0582 & 0.0397 & 0.0397 \\
0.0566 & 0.156  & 0.155  & 0.0970 & 0.0969 & 0.0597 & 0.0597 \\
0.0713 & 0.275  & 0.269  & 0.179  & 0.178  & 0.121  & 0.121  \\
0.0897 & 0.461  & 0.456  & 0.270  & 0.273  & 0.182  & 0.182  \\
0.113  & 0.639  & 0.624  & 0.387  & 0.387  & 0.248  & 0.249  \\
0.142  & 0.942  & 0.921  & 0.621  & 0.611  & 0.406  & 0.403  \\
0.179  & 1.17   & 1.14   & 0.857  & 0.845  & 0.593  & 0.615  \\
0.225  & 1.72   & 1.63   & 1.15   & 1.16   & 0.821  & 0.792  \\
0.284  & 1.53   & 1.33   & 1.39   & 1.30   & 1.17   & 1.16   \\
0.357  & 1.80   & 1.62   & 1.37   & 1.22   & 1.36   & 1.25   \\
0.450  &        &        & 1.67   & 1.52   & 1.26   & 1.17   \\
0.566  &        &        &        &        & 1.49   & 1.36   \\
\hline
\end{tabular*}
\end{center}
\end{minipage}
\end{table*}
\begin{table*}
\begin{minipage}{256pt}
\begin{center}
\caption {CHDM recovered linear power spectra
$\Delta^2(k)=\frac{k^3}{2\pi^2}P(k)$}
\label{chdmfam_table}
\begin{tabular*}{256pt}{ccccc}
& \multicolumn{2}{c}{$b_{\rm I}=0.8$}
& \multicolumn{2}{c}{$b_{\rm I}=1.0$} \\
$k$ (\hmpcinv) & CHDM$_{0.7}$ & CHDM$_{0.8}$ & CHDM$_{0.7}$ &
CHDM$_{0.8}$ \\
\hline
0.0142 & 0.0016 & 0.0016 & 0.0010 & 0.0010 \\
0.0179 & 0.0018 & 0.0018 & 0.0012 & 0.0012 \\
0.0225 & 0.0044 & 0.0044 & 0.0030 & 0.0030 \\
0.0284 & 0.0122 & 0.0122 & 0.0085 & 0.0085 \\
0.0357 & 0.0336 & 0.0336 & 0.0237 & 0.0237 \\
0.0450 & 0.0500 & 0.0499 & 0.0350 & 0.0350 \\
0.0566 & 0.0863 & 0.0861 & 0.0530 & 0.0529 \\
0.0713 & 0.159  & 0.158  & 0.111  & 0.111  \\
0.0897 & 0.236  & 0.236  & 0.170  & 0.169  \\
0.113  & 0.341  & 0.350  & 0.235  & 0.233  \\
0.142  & 0.537  & 0.615  & 0.368  & 0.388  \\
0.179  & 0.647  & 0.813  & 0.540  & 0.619  \\
0.225  & 0.768  & 1.11   & 0.666  & 0.854  \\
0.284  & 0.797  & 1.15   & 0.768  & 1.11   \\
0.357  & 0.762  & 1.06   & 0.690  & 0.868  \\
0.450  & 0.830  & 1.26   & 0.750  & 1.03   \\
0.566  &        &        & 0.795  & 1.16   \\
\hline
\end{tabular*}
\end{center}
\end{minipage}
\end{table*}
\begin{table}
\caption {SCDM recovered linear power spectra
\label{scdmfam_table}
$\Delta^2(k)=\frac{k^3}{2\pi^2}P(k)$}
\begin{center}
\begin{tabular}{cccc}
$k$ (\hmpcinv) & $b_{\rm I}=0.6$ & $b_{\rm I}=0.8$ & $b_{\rm I}=1.0$ \\
\hline
0.0142 & 0.0028 & 0.0016 & 0.0010 \\
0.0179 & 0.0028 & 0.0018 & 0.0012 \\
0.0225 & 0.0071 & 0.0044 & 0.0030 \\
0.0284 & 0.0189 & 0.0122 & 0.0085 \\
0.0357 & 0.0513 & 0.0335 & 0.0236 \\
0.0450 & 0.0765 & 0.0498 & 0.0349 \\
0.0566 & 0.135  & 0.0861 & 0.0527 \\
0.0713 & 0.237  & 0.159  & 0.111  \\
0.0897 & 0.412  & 0.243  & 0.173  \\
0.113  & 0.664  & 0.379  & 0.242  \\
0.142  & 1.25   & 0.756  & 0.488  \\
0.179  & 1.69   & 1.10   & 0.807  \\
0.225  & 2.10   & 1.70   & 1.15   \\
0.284  & 1.65   & 1.75   & 1.70   \\
0.357  & 2.10   & 1.50   & 1.11   \\
0.450  &        & 1.94   & 1.42   \\
0.566  &        &        & 1.71   \\
\hline
\end{tabular}
\end{center}
\end{table}

They contain the linearized observational power
spectra for the \lcdm, CHDM, and SCDM models, respectively.

\section{Conclusions}
\label{conclusions}

We have reconstructed the linear power spectra from observations for
several current cosmological models based on rigorously tested
linearization methods.  We have found that:

(i) The linearized power spectra of all \lcdm\ and CHDM models are
systematically below their respective linear power spectra using the
bias ratio in Eq.\ (\ref{biasratio}) normalized so that the bias of
\iras\ galaxies $b_{\rm I}=1.0$.  It was necessary to adjust the
normalization of the bias ratio so that $b_{\rm I}<1.0$ to obtain the best
fit.  While this implies that optically selected galaxies must be
antibiased ($b_{\rm O}<1$) to be consistent with models, this may be
acceptable since a recent \potent\ analysis (Hudson \etal\ 1995) allows a
slight antibias for some \lcdm\ models.

(ii) The linearization of the SCDM model cannot be made to fit its
linear power spectrum under any circumstances and fits very poorly at
best.  This simply reconfirms the incompatibility of SCDM with
observations, as has long been recognized.

(iii) The \lcdm\ models all appear moderately successful, as indicated
by our `$P$ ratio' indicator in Table~\ref{fitsum_table}, but share
difficulties in matching the shape of their linear spectra.  This is a
consequence of our finding that the linearized spectra of all \lcdm\
models share a common shape for a given bias.  The only exception is
the T\lcdm$_{0.4}$ model, though it deviates from the others by at
most 10 per cent at $k\sim 0.5$ \hmpcinv.  This is to be expected since
T\lcdm$_{0.4}$ has more curvature by design.  The best fitting \lcdm\
model is \lcdm$_{0.5}$ as determined from Table~\ref{fitsum_table}.

(iv) The CHDM models seem to do an excellent job of recovering their
linear power spectra from observations.  They require relatively small
$b_{\rm I}$ to fit well and have the lowest $P$ ratios.  Ma (1996) found that
a slight tilt of $n=0.9$--$0.95$ is needed to bring $h=0.5$ CHDM models into
agreement with the PD94 reconstructed spectrum.  However, if one reconstructs
the linear spectrum self-consistently, using the nonlinear spectrum for the
CHDM models considered here, there is good agreement without any tilt.

(v) There are physical motivations to include a scale-dependent bias
in the treatment of observational power spectra. Such a bias would
increase with higher $k$ and would make the reconstructions at $k\ga
0.2$ \hmpcinv\ worse than they already are.  A bias that
decreased with scale could significantly improve the agreement between
the assumed and reconstructed power spectrum for each model,
especially the \lcdm\ models, but this appears to be an entirely ad
hoc assumption.

(vi) There is a unique model-independent linear power spectrum that
can be recovered from observations in the region $0.01\la k\la
0.1$ \hmpcinv, at least for the models considered here.

We have demonstrated that the linearization method is accurate to
within 20 per cent up to linear scales of at least $k=1$ \hmpcinv\ and
possibly higher.  Because the most favorable models are nearly
indentical up to the highest reliable bin, a power spectrum with small
errors that extends from large nonlinear scales to $k\sim
10$ \hmpcinv\ or more would be very helpful in constraining the
shape of theoretical linear power spectra.

\section*{Acknowledgements}

Simulations were performed on the CM-5 and Convex 3880 at the National
Center for Supercomputing Applications, University of Illinois,
Champaign-Urbana, IL, and the SP2 at the Theory Center, Cornell
University, Ithaca, NY.  We acknowledge support from NSF and NASA
grants at NMSU and UCSC.

\begin{table*}
\begin{minipage}{348pt}
\begin{center}
\caption {Linear Power Spectra for Cold Matter at $z=0$}
\label{tbl:cold_spectra}
\begin{tabular*}{348pt}{lccccccc}
Model & $A$ ($h^{-3-n}$ Mpc$^{3+n}$)& $n$ & $a_1$ & $a_2$ & $a_3$ &
 $a_4$ & $b$\\
\hline
\lcdm$_{0.3}$ &$4.507\times 10^6$&  1&-1.3050&33.590&68.973&157.74&1.8606\\
\lcdm$_{0.4}$ &$2.663\times 10^6$&  1&-1.2120&29.224&57.044&119.53&1.8623\\
T\lcdm$_{0.4}$&$1.697\times 10^6$&0.9&-1.2120&29.224&57.044&119.53&1.8623\\
\lcdm$_{0.5}$ &$9.880\times 10^5$&  1&-1.0752&24.087&36.390&78.732&1.8533\\
\\
CHDM$_{0.7}$  &$5.960\times 10^5$&  1&-1.0421&19.270&-56.526&310.68&1.8270\\
CHDM$_{0.8}$  &$7.827\times 10^5$&  1&-0.1828&0.4526& 73.673&104.66&1.8571\\
\hline
\end{tabular*}
\end{center}
\end{minipage}
\end{table*}

\begin{table*}
\begin{minipage}{260pt}
\begin{center}
\caption {Linear Power Spectra for Hot Matter at $z=0$}
\label{tbl:hot_spectra}
\begin{tabular*}{260pt}{lccccc}
Model & $c$ & $d_1$ & $d_2$ & $d_3$ & $d_4$\\
\hline
CHDM$_{0.7}$&0.65837&-0.51390&0.34412&0.032050&0.023225\\
CHDM$_{0.8}$&0.76462&-0.71510&1.6616&-1.0933&0.8900\\
\hline
\end{tabular*}
\end{center}
\end{minipage}
\end{table*}

\appendix
\section{Fitting functions for linear power spectra}
\label{app:fits}
\subsection{Calculations}

The linear calculations were done using the techniques of Holtzman (1989).
Perturbations in radiation, baryons, cold dark matter, and massive and
massless neutrinos were caluclated from a scale factor of $1\times 10^{-10}$
to the present (where the scale factor is $a\equiv 1/(1+z)$). For models
without massive neutrinos, three massless neutrino species were used;
for models with massive neutrinos, the mass was either placed in a 
single neutrino species or two equal-mass species, with the other species
remaining massless.

For the
radiation, the angular dependence of the perturbations was handled by
expanding the perturbation in Legendre polynomials, $\delta_{\rm r} = \sum 
\Delta_l(k,a) P_l (cos \theta)$. The calculation started with a small
number of orders, but additional orders were added as they became needed
to keep the final $P(k)$ results accurate to 1 per cent.
For large scales ($k<0.7$ \hmpcinv), the radiation perturbations were
integrated all the way to the present. For smaller scales, they were 
integrated through recombination, and then the analytic solution presented
by Bond and Efstathiou (1987) was used to calculate the perturbations
at the present time; no approximations were made for any scale for the
radiation perturbations. The recombination history of Peebles (1967) was
used for hydrogen, but a helium mass fraction of 0.25 was included.

For the massive neutrinos, the perturbations are a function of both angle
and neutrino momentum. We calculated the evolution for 15 separate 
momenta up to the time when the neutrinos became nonrelativistic; this
was done using the integro-differential equations presented by Bond
and Szalay (1983). This technique allows only the 0th and 1st orders 
of the neutrino angular distribution to be computed without following
higher orders; these are the only orders than enter as the graviational 
source terms. For small scales, even this computation can get
very expensive, so the massive neutrino perturbations were manually damped
when they became gravitationally negligible (which occurs because the
perturbations are destroyed by free-streaming).  When the neutrinos
became nonrelativistic, the full angular distribution for each momenta
was calculated,
and the subsequent evolution was computed using the full set of coupled
differential equations exactly analagous to those used for the radiation
perturbation.

\subsection{Fitting functions}

For convenience, we provide fits for the power spectrum of the hot and
cold components as a function of wavenumber, $k$.  For all of the calculations
presented in this paper, we used the computed spectra directly, rather than the
fits, to obtain results.  The fits are good to better than 5 per cent in total
power, in the worst case.
Errors of 1--2 per cent are more typical for $k=0.1$--$30$ \hmpcinv.
For fitting functions, we used
\begin{equation}
P_{\rm c+b}(k)=\frac{Ak^n}{\left(1+a_1k^{1/2}+a_2k+a_3k^{3/2}+a_4k^2\right)^b}
\label{eqn:pcold}
\end{equation}
for the cold and baryonic matter, and
\begin{equation}
P_\nu(k)=\frac{P_{\rm c+b}(k)\exp(-ck^{1/2})}
          {\left(1+d_1k^{1/2}+d_2k+d_3k^{3/2}+d_4k^2\right)^2}
\label{eqn:phot}
\end{equation}
for the massive neutrinos.  The total power is given by
\begin{equation}
P(k)=\left[\frac{\Omega_{\rm b}+\Omega_{\rm c}}{\Omega_0}\sqrt{P_{\rm c+b}(k)}+
             \frac{\Omega_\nu}{\Omega_0}\sqrt{P_\nu}(k)\right]^2.
\end{equation}

The parameters for all the models considered in this paper
are presented in Table~\ref{tbl:cold_spectra}
for the cold component and in Table~\ref{tbl:hot_spectra}
for the hot component; the coefficients assume
the use of units of \hmpcinv\ for k in equations~\ref{eqn:pcold} and
\ref{eqn:phot}.  The normalization $A$ is set to reproduce the $\sigma_8$ used
in this paper.


\section*{References}

\def\hangin{\par\hangindent .5cm\hangafter=1\noindent}

\hangin
Bardeen, J., Bond, J.~R., Kaiser, N., \& Szalay, A.~S.,
 1986, ApJ, 304, 15 (BBKS)

\hangin
Baugh, C.~M., \& Efstathiou, G., 1993, MNRAS, 265, 145

\hangin
Borgani, S., Moscardini, L., Plionis, M., Gorski, K.,
 Holtzman, J., Klypin, A., Primack, J.R.,  Smith, C.L. \&
 Stompor, R. 1997, New Astronomy, in press; preprint astro-ph/9611100

\hangin
Carroll, S.~M., Press, W.~H., \& Turner, E.~L., 1992,
 ARAA, 30, 499

\hangin
Coble, K., Dodelson, S., \& Frieman, J., 1996,
 preprint astr-ph/9608122

\hangin
Feldman, H.~A., Kaiser, N., \& Peacock, J.~A., 1994,
 ApJ, 426, 23

\hangin
Hamilton, A. J.~S., Kumar, P., Lu, E., \& Matthews, A., 1991,
 ApJ, 374, L1 (HKLM)

\hangin
Hockney, R.~W., \& Eastwood, J.~W., 1981,
 {\em Numerical simulations using particles},
 McGraw-Hill, New York

\hangin
Holtzman, J., 1989, ApJS, 71, 1

\hangin
Hu, W., \& Sugiyama, N. 1996, ApJ, 471, 542

\hangin
Hudson, M.~J., Dekel, A., Courteau, S., Faber, S.~M., \&
 Willick, J.~A., 1995, MNRAS, 274, 305

\hangin
Jain, B., Mo, H.~J., \& White, S. D.~M., 1995,
 MNRAS, 276, L25

\hangin
Kaiser, N., 1987, MNRAS, 227, 1

\hangin
Kates, R.~E., Kotok, E.~V., \& Klypin, A., 1991,
 A\&A, 243, 295

\hangin
Kauffmann, G., Nusser, A., \& Steinmetz, M. 1995, preprint astro-ph/9512009

\hangin
Klypin, A., Holtzman, J., Primack, J., \& Regos, E., 1993,
 ApJ, 416, 1

\hangin
Klypin, A., Primack, J.R., \& Holtzman, J. 1996, ApJ, 466, 1 (KPH96)

\hangin
Liddle, A.~R., Lyth, D.~H., Schaefer, R.~K., Shafi, Q., \& Viana, P. T.~P.,
  1996a, MNRAS, 281, 531

\hangin
Liddle, A.~R., Lyth, D.~H., Viana, P. T.~P., \& White, M.,
 1996b, MNRAS, 282, 281

\hangin
Lahav, O., Lilje, P., Primack, J.R., \& Rees, M. 1991, MNRAS,
 251, 128

\hangin
Loveday, J., Efstathiou, G., Peterson, B.~A., \& Maddox, S.~J., 1992,
 ApJ, 400, L43

\hangin
Ma, C.~P., 1996, ApJ, 471, 13

\hangin
Mo, H.~J., Peacock, J.~A., \& Xia, X.~Y., 1993,
 MNRAS, 260, 121

\hangin
Peacock, J.~A., 1992,
 in V. Martinez, M. Portilla, \& D. Saez (eds.), {\em New insights
 into the Universe}, Proc. Valencia summer school,
 (Springer, Berlin), p.~1

\hangin
Peacock, J.~A. 1996, preprint astro-ph/9608151

\hangin
Peacock, J.~A., \& Dodds, S.~J., 1994, MNRAS, 267, 1020 (PD94)

\hangin
Peacock, J.~A., \& Dodds, S.~J., 1996, MNRAS, 280, L19 (PD96)

\hangin
Peacock, J.~A., \& Nicholson, D., 1991, MNRAS, 253, 307

\hangin
Peacock, J.~A. \& West, M.~J., 1992, MNRAS, 259, 494

\hangin
Peebles, P. J.~E., 1980,
 {\em The Large-Scale Structure of the Universe}
 (Princeton Univ. Press, Princeton, NJ)

\hangin
Primack, J.R., Holtzman, J., Klypin, A., \& Caldwell, D.O.
 1995, Phys. Rev. Lett., 74, 2160

\hangin
Smith, C., 1995,
 {\em Master's thesis}, New Mexico State University

\hangin
Sugiyama, N., 1995, ApJS 100, 281

\hangin
Vogeley, M.~S., Park, C., Geller, M.~J., \& Huchra, J.~P., 1992,
 ApJ, 391, L5

\hangin
Yepes, G., Kates, R., Khokhlov, A., \& Klypin, A. 1997, MNRAS, 284, 235

\ifref
\pagestyle{empty}
\thispagestyle{empty}
\renewcommand{\thefigure}{{\arabic{figure}}}
\renewcommand{\thetable}{{\arabic{table}}}

\processdelayedfloats
\begin{figure*}
  \resizebox{\textwidth}{!}{\includegraphics{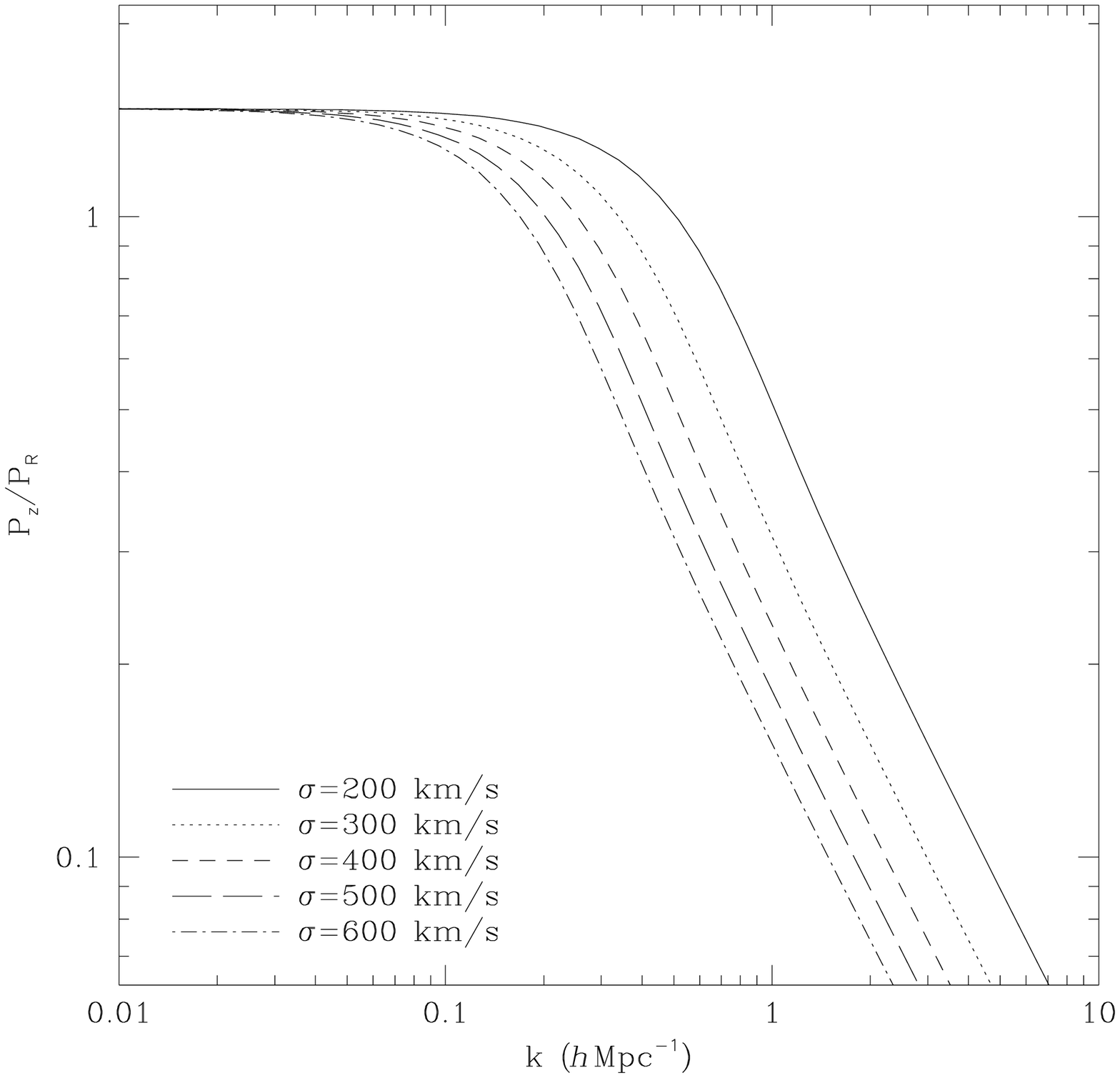}}
\end{figure*}

\processdelayedfloats
\begin{figure*}
  \resizebox{\textwidth}{!}{\includegraphics{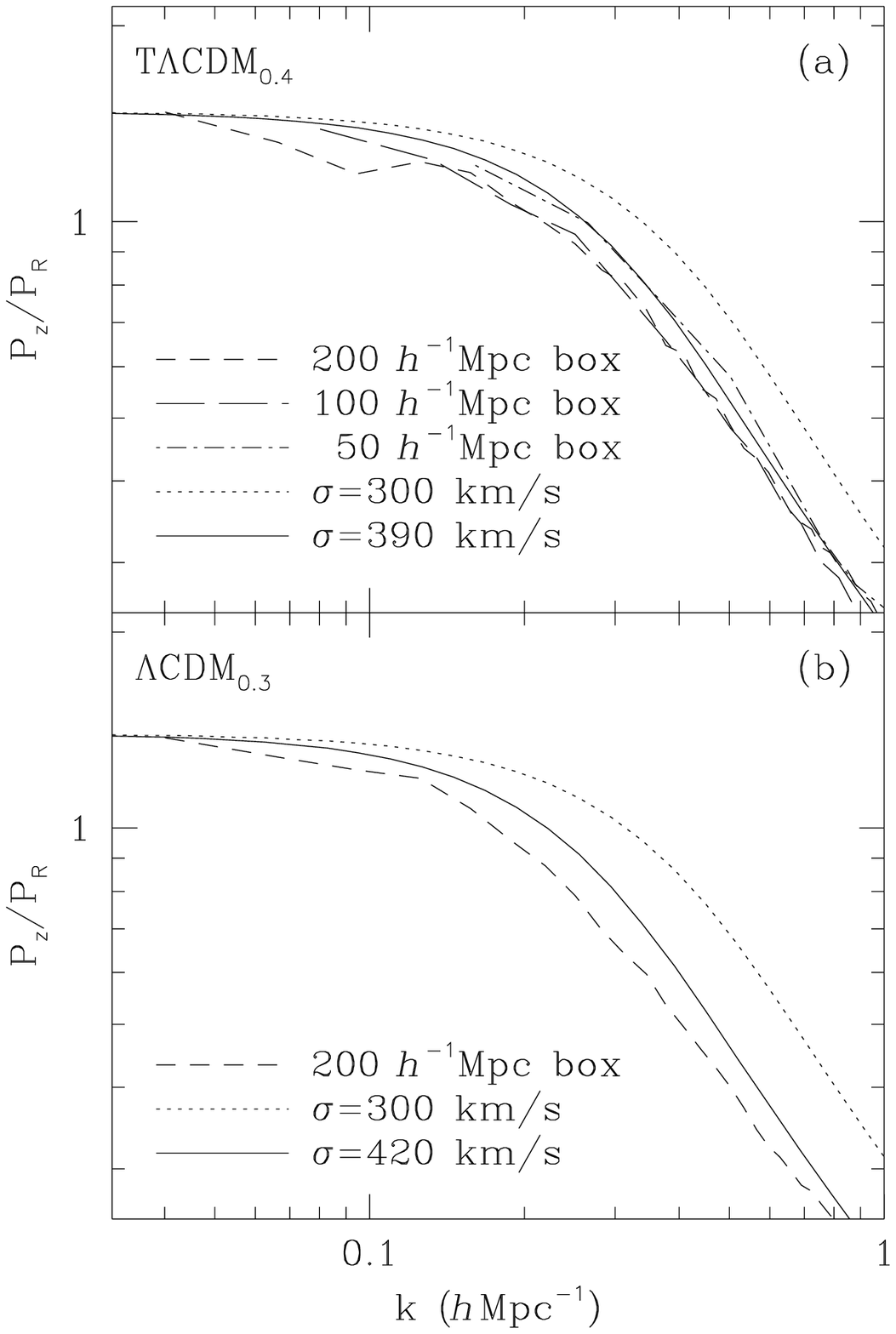}}
\end{figure*}

\processdelayedfloats
\begin{figure*}
  \resizebox{\textwidth}{!}{\includegraphics{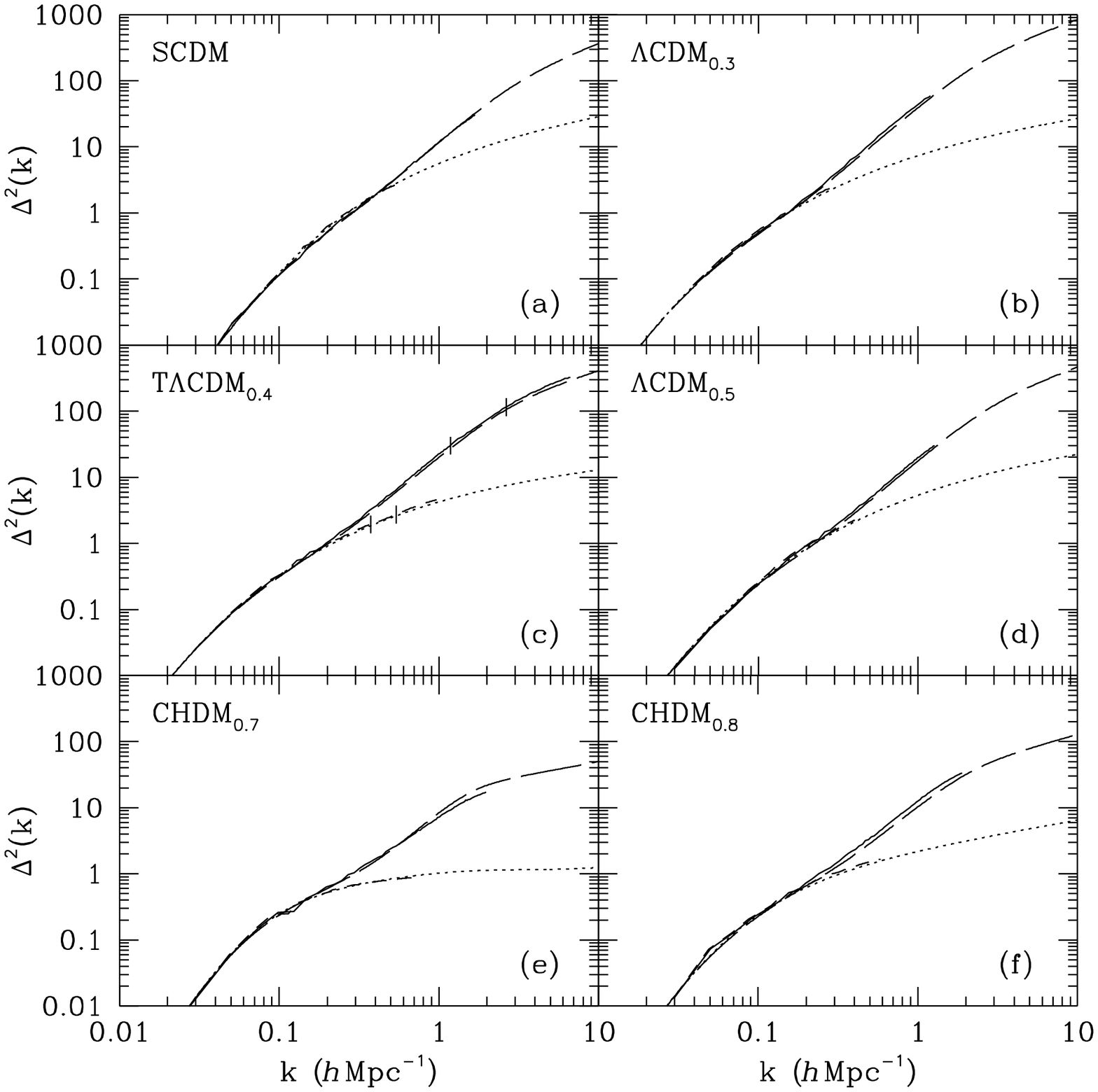}}
\end{figure*}

\processdelayedfloats
\begin{figure*}
  \resizebox{\textwidth}{!}{\includegraphics{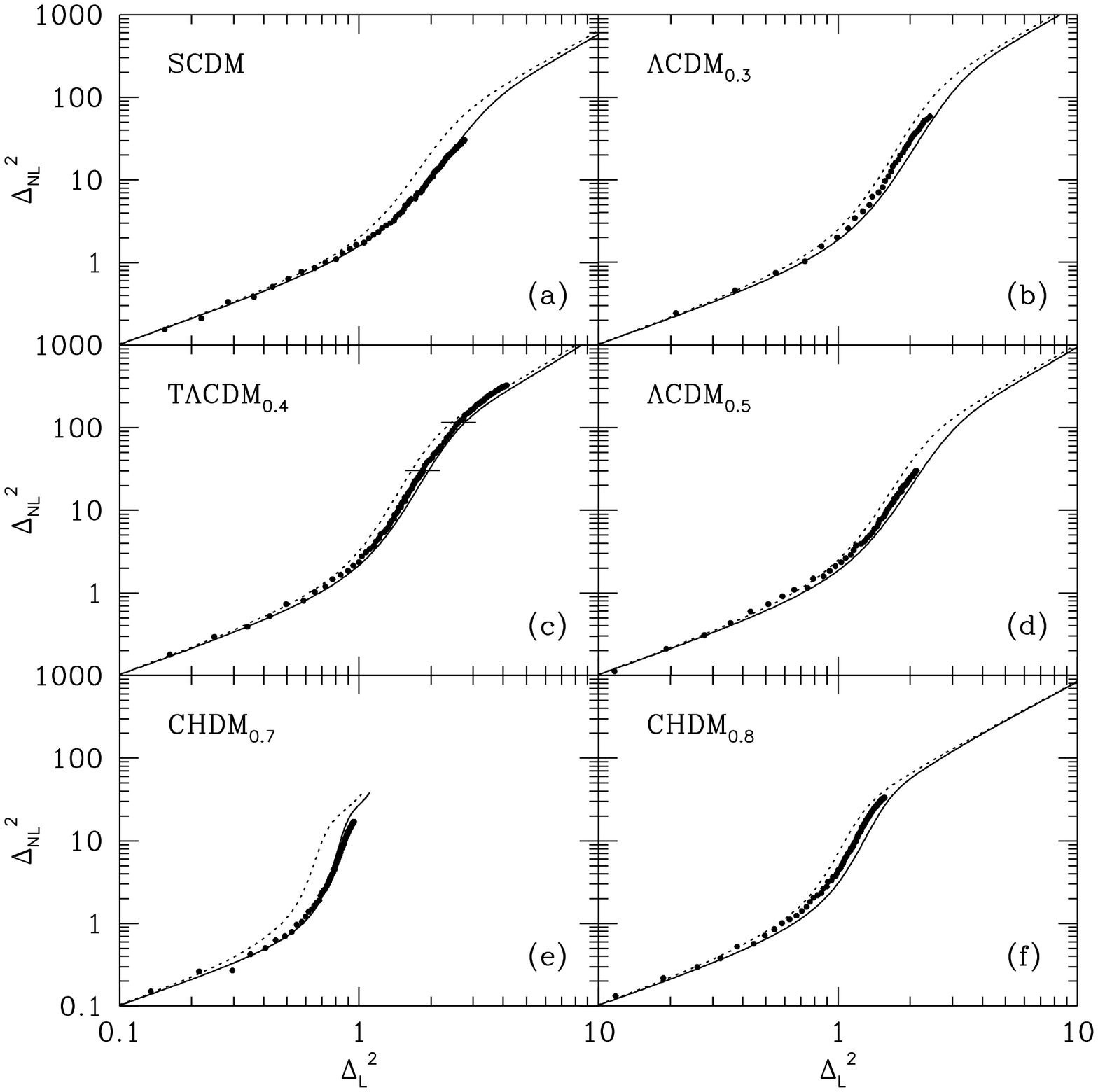}}
\end{figure*}

\processdelayedfloats
\begin{figure*}
  \resizebox{\textwidth}{!}{\includegraphics{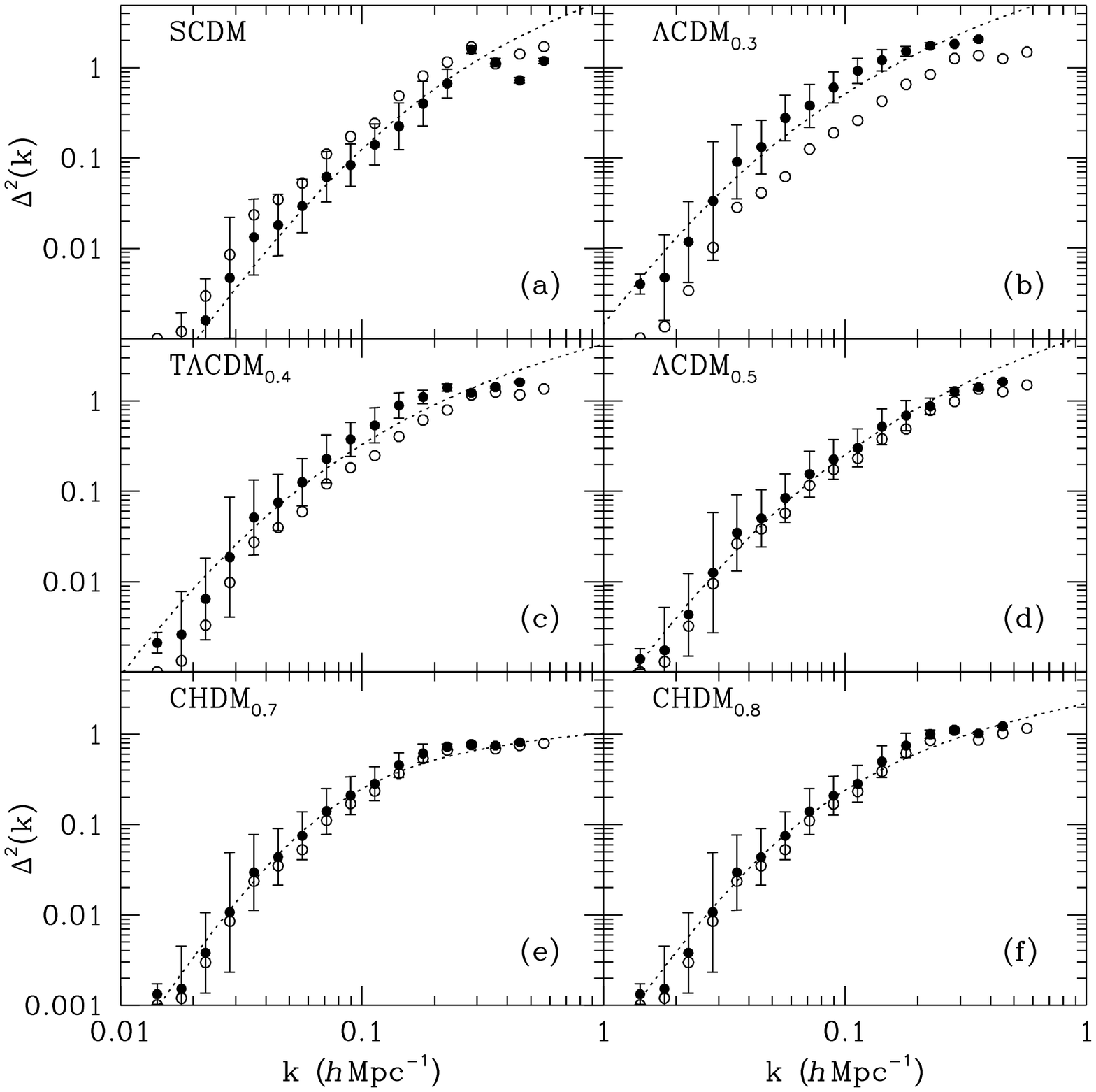}}
\end{figure*}

\processdelayedfloats
\begin{figure*}
  \resizebox{\textwidth}{!}{\includegraphics{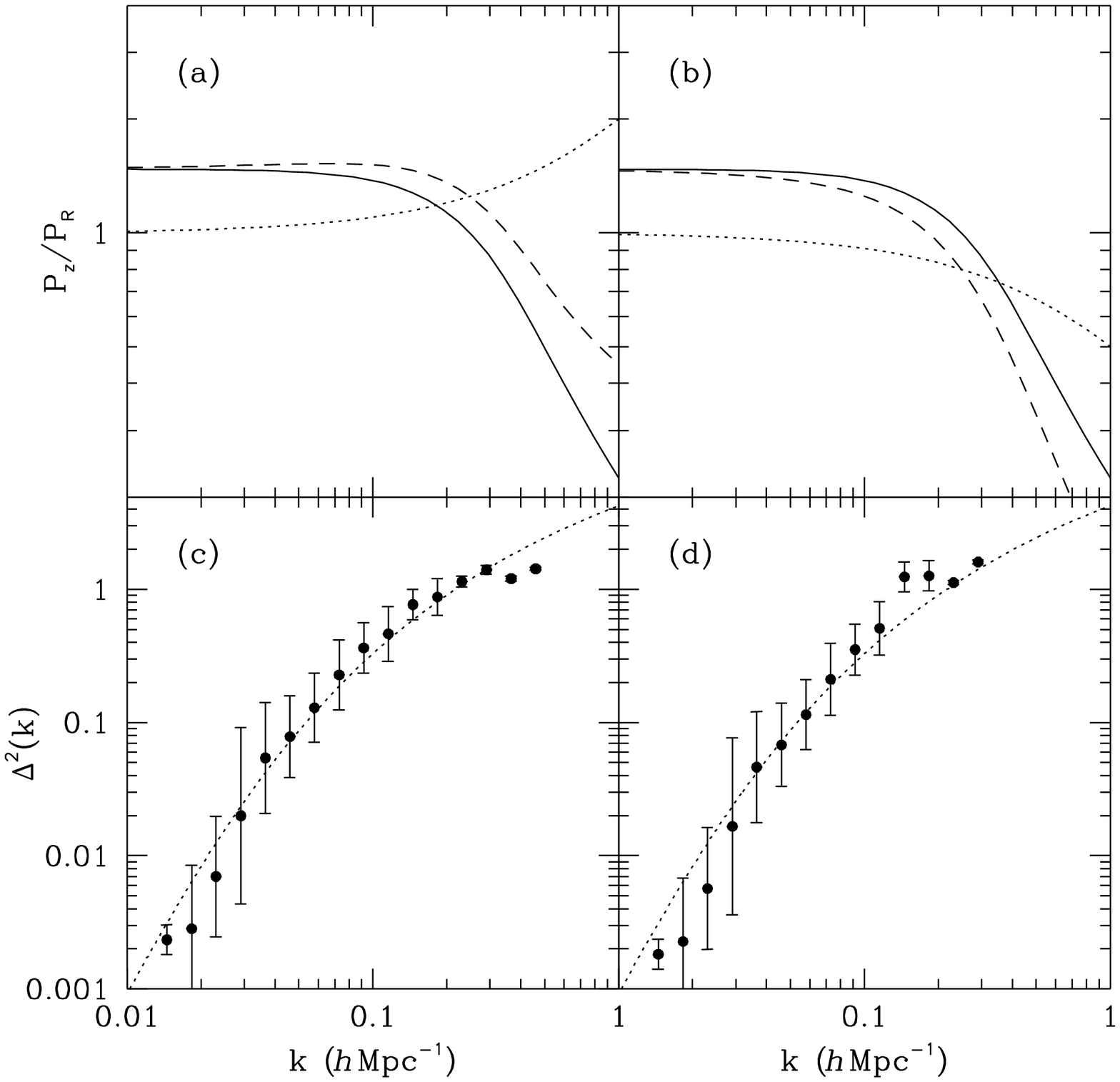}}
\end{figure*}

\processdelayedfloats
\begin{figure*}
  \resizebox{\textwidth}{!}{\includegraphics{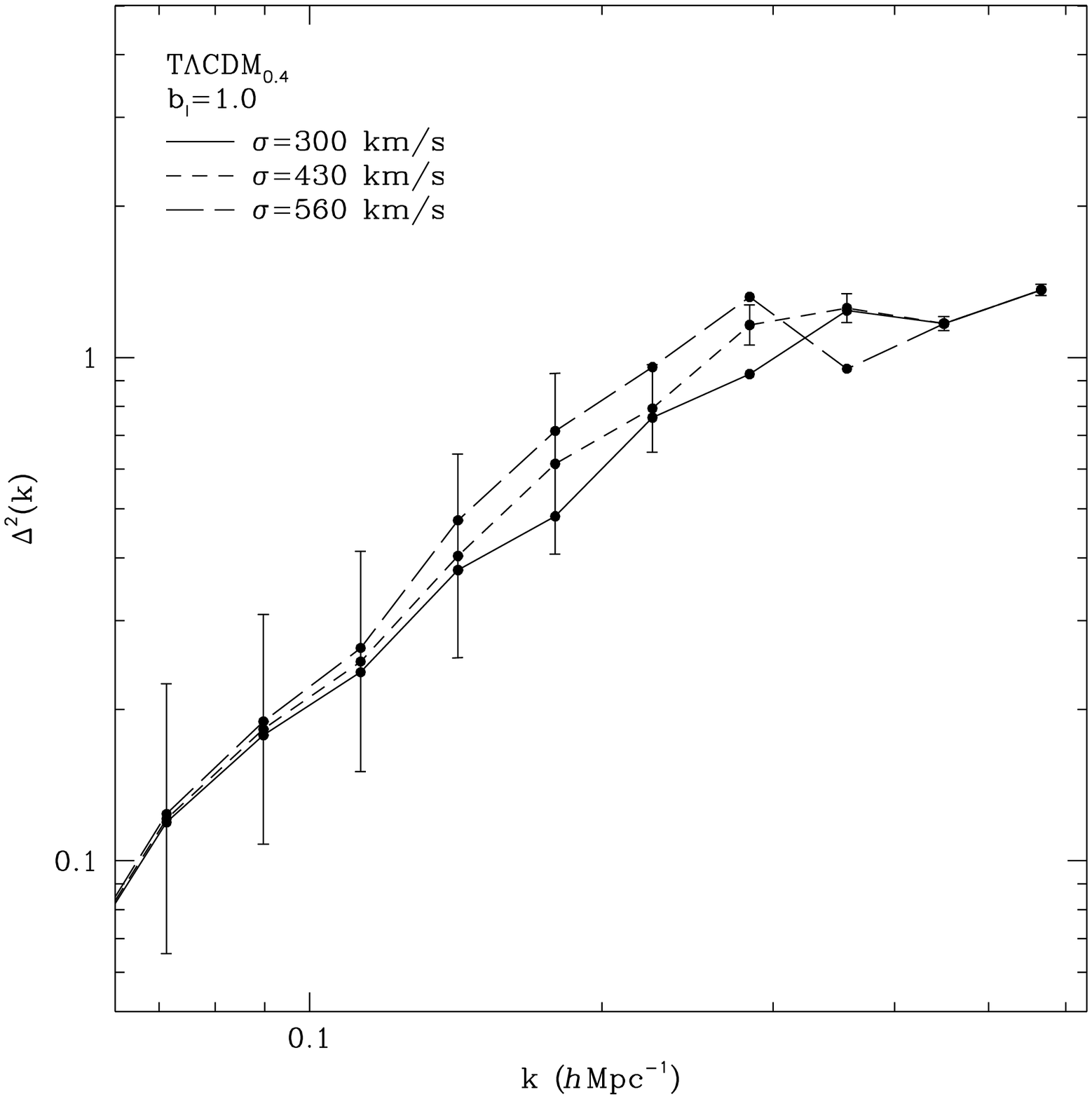}}
\end{figure*}

\processdelayedfloats
\begin{figure*}
  \resizebox{\textwidth}{!}{\includegraphics{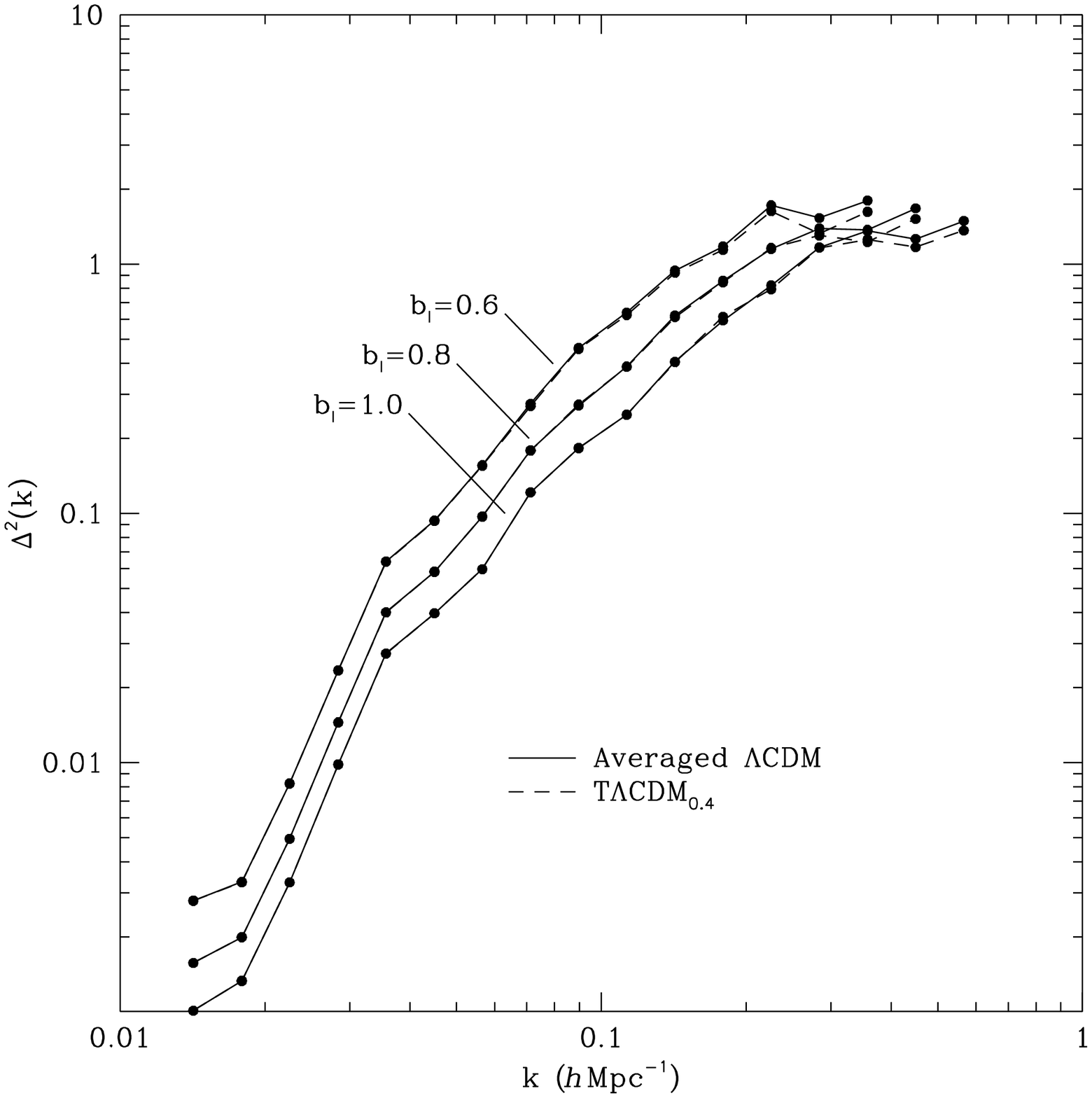}}
\end{figure*}

\processdelayedfloats
\begin{figure*}
  \resizebox{\textwidth}{!}{\includegraphics{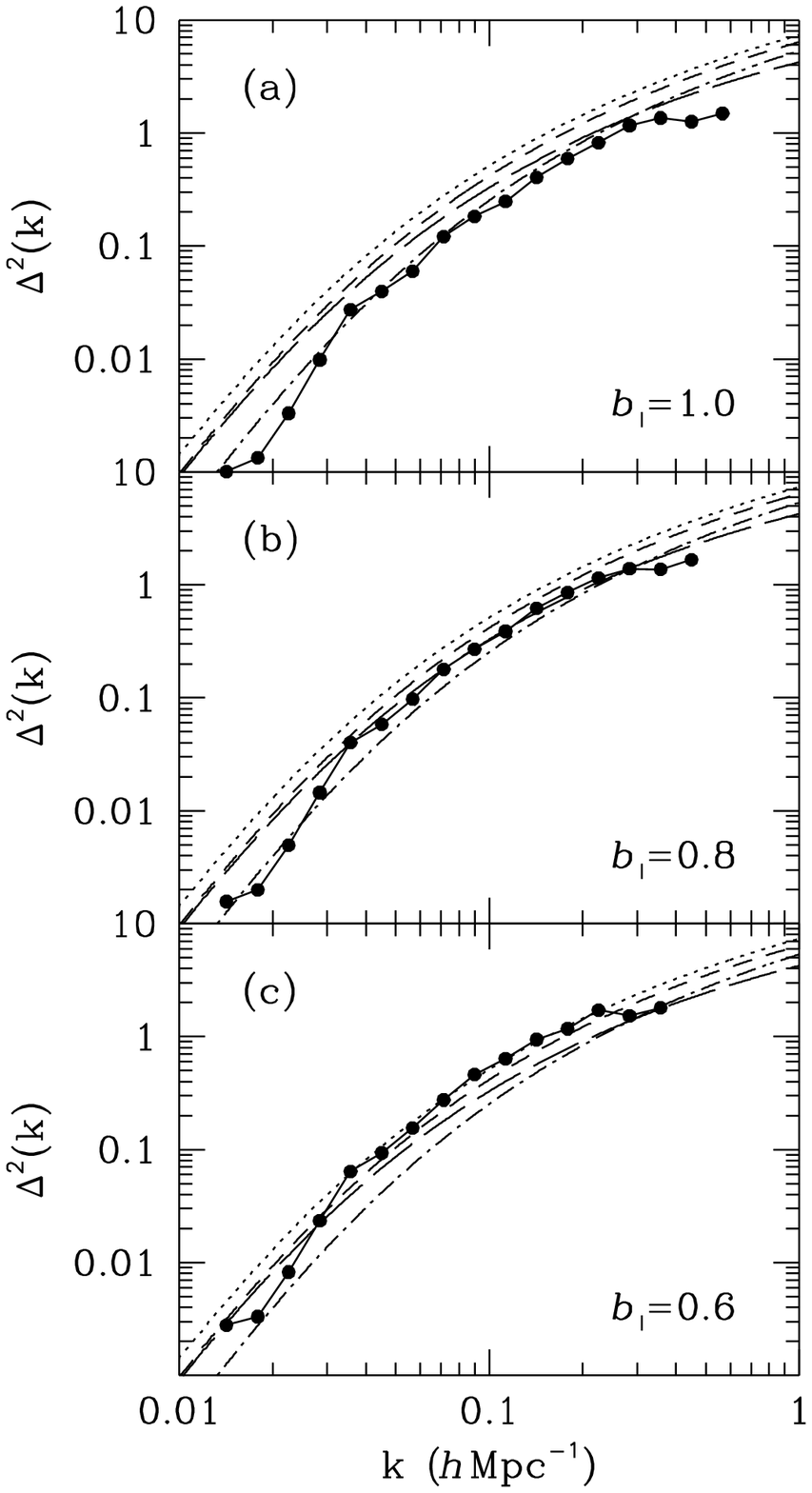}}
\end{figure*}

\processdelayedfloats
\begin{figure*}
  \resizebox{\textwidth}{!}{\includegraphics{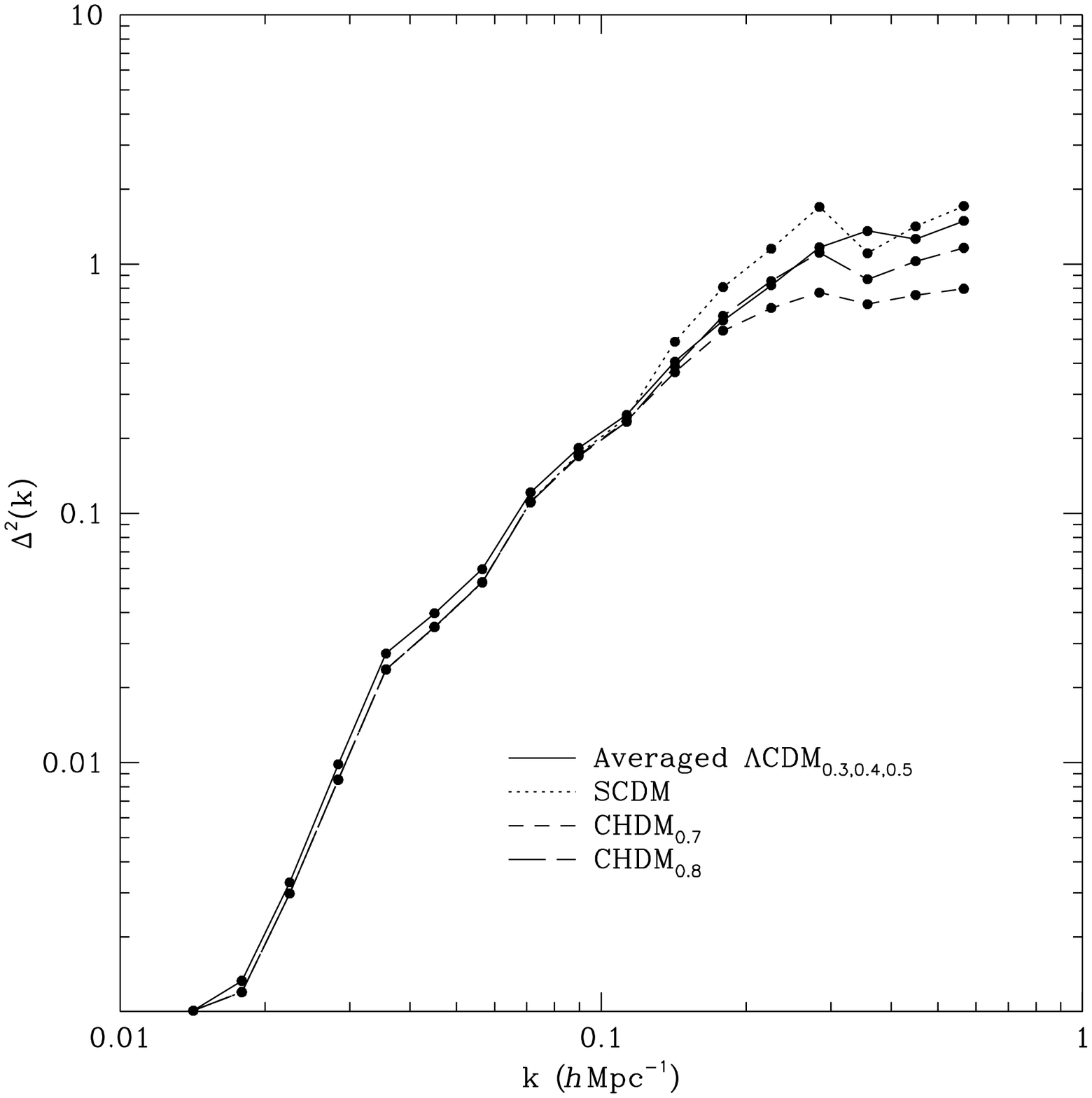}}
\end{figure*}

\fi

\end{document}